\documentclass[twocolumn, 10pt]{IEEEtran}

\usepackage{graphicx}
\usepackage{amssymb}
\usepackage{amsmath}
\usepackage{mathtools}
\usepackage{cite}
\usepackage{stfloats}
\usepackage{subfigure}
\usepackage{psfrag}
\usepackage[mathscr]{euscript}
\usepackage{acronym}  
\usepackage{booktabs} 
\usepackage[table]{xcolor}
\usepackage{hyperref} 
\makeatletter \def\Hy@BorderArrayPatch{} \makeatother

\acrodef{MEC}{mobile edge computing}
\acrodef{AP}{access point}
\acrodef{PPP}{Poisson point process}
\acrodefplural{PPP}[PPPs]{Poisson point processes}
\acrodef{pdf}{probability density function}
\acrodef{cdf}{cumulative density function}
\acrodef{HetNet}{heterogeneous network}
\acrodef{SINR}{signal-to-interference-plus-noise ratio}
\acrodef{SIR}{signal-to-interference ratio}
\acrodef{SECP}{successful edge computing probability}
\acrodef{SCP}{successful computation probability}
\usepackage{color}
\usepackage{dsfont}
\usepackage{bbm}

\newtheorem{theorem}{Theorem}
\newtheorem{lemma}{Lemma}
\newtheorem{corollary}{Corollary}

\def \MS{\text{m}}
\def \US{\text{u}}

\def \OFFL{\text{o}}
\def \SEC{\text{s}}
\def \COMP{\text{cp}}
\def \COMM{\text{cm}}

\newcommand{\pui}[1]{p_{\US, #1}}
\newcommand{\pmk}[1]{p_{\MS, #1}}
\newcommand{\poik}[2]{p_{\OFFL, #1, #2}}

\newcommand{\pec}{p_{\SEC}}
\newcommand{\pech}{\hat{p}_{\SEC}}
\newcommand{\pecik}[2]{p_{\SEC,#1,#2}}
\newcommand{\pecikh}[2]{\hat{p}_{\SEC,#1,#2}}

\newcommand{\pcomp}{p_{\COMP}}

\newcommand{\pcpik}[2]{p_{\COMP,#1,#2}}
\newcommand{\pcp}{p_{\COMP}}
\newcommand{\pcmuik}[2]{p_{\COMM, #1, #2}^{\left(u\right)}}
\newcommand{\pcmdik}[2]{p_{\COMM, #1, #2}^{\left(d\right)}}
\newcommand{\pcmqik}[2]{p_{\COMM, #1, #2}^{\left(q\right)}}
\newcommand{\px}[1] {{\mathbb{P}}\left\{{#1}\right\}}
\newcommand{\fsvik}[2]{f_{T_{\text{sv},#1,#2}}\left(t\right)}
\newcommand{\fsvikr}[2]{f_{T_{\text{sv},#1,#2}}\left(r\right)}
\newcommand{\fsvk}[1]{f_{T_{\text{sv},#1}}\left(t\right)}

\newcommand{\lbdm}{\lambda_{\MS}}
\newcommand{\lbdu}{\lambda_{\US}}
\newcommand{\lbdmk}[1]{\lambda_{\MS,#1}}
\newcommand{\lbdui}[1]{\lambda_{\US,#1}}
\newcommand{\lbdaik}[2]{\nu_{#1,#2}}
\newcommand{\lbdak}[1]{\nu_{#1}}

\newcommand{\muk}[1]{\mu_{#1}}
\newcommand{\dit}[2]{D_{\text{u}#1}}
\newcommand{\betauik}[2]{\beta_{#1,#2}^{\left(u\right)}}
\newcommand{\betadik}[2]{\beta_{#1,#2}^{\left(d\right)}}
\newcommand{\betawik}[2]{\beta_{#1,#2}^{\left(q\right)}}

\newcommand{\drwik}[2]{R_{#1,#2}^{\left(q\right)}}
\newcommand{\ncck}[1]{\mathcal{NC}_{#1}}

\def \PPP{\mathsf{\bold{\Phi}}}
\newcommand{\pppm}{\PPP_{\MS}}
\newcommand{\pppmk}[1]{\PPP_{\MS, #1}}
\newcommand{\pppu}{\PPP_{\US}}
\newcommand{\pppui}[1]{\PPP_{\US, #1}}

\def \BIAS{B}
\newcommand{\biasik}[2]{\BIAS_{\text{u}#1\triangleright\text{s}#2}}
\newcommand{\biasikh}[2]{\hat{\BIAS}_{\text{u}#1\triangleright\text{s}#2}}

\newcommand{\biasoptik}[2]{\BIAS^{*}_{\text{u}#1\triangleright\text{s}#2}}
\def \WEIGHT{\mathcal{W}}
\newcommand{\weightik}[2]{\WEIGHT_{\text{u}#1\triangleright\text{s}#2}}
\newcommand{\weightikh}[2]{\hat{\WEIGHT}_{\text{u}#1\triangleright\text{s}#2}}

\def \PWR{P}
\newcommand{\pwrui}[1]{\PWR_{\US,#1}}
\newcommand{\pwrmk}[1]{\PWR_{\MS,#1}}
\newcommand{\pwrmkh}[1]{\hat{\PWR}_{\MS,#1}}
\def \BW{W}
\newcommand{\bw}[1]{\BW^{#1}}
\def \Z{Z}
\def \x{\bold{X}}

\def \lx{\bold{x}}
\def \ly{\bold{y}}
\newcommand{\zik}[2]{\Z_{#1,#2}}
\def\SetK{\mathcal{K}}
\def\SetI{\mathcal{I}}
\def\SetQ{\mathcal{Q}}

\newcommand{\punit}{U_{\text{s}}}
\newcommand{\rhok}[1]{\rho_{#1}}
\newcommand{\ttg}{T_{\text{t}}}
\newcommand{\ttgi}[1]{T_{\text{t},#1}}

\newcommand{\tcik}[2]{T_{\text{c},#1,#2}}
\newcommand{\ts}{T_{\text{s}}}
\newcommand{\tti}[1]{T_{\text{total},#1}}

\newcommand{\tquk}[1]{T_{\text{w},#1}}

\newcommand{\tsvik}[2]{T_{\text{sv},#1,#2}}
\newcommand{\tsvk}[1]{T_{\text{sv},#1}}
\newcommand{\tsvkl}[1]{\hat{\tau}_{\text{sv},k}\left(s\right)}
\newcommand{\tcmuik}[2]{T_{\COMM,#1,#2}^{(\text{u})}}
\newcommand{\tcmdik}[2]{T_{\COMM,#1,#2}^{(\text{d})}}
\newcommand{\tcmwik}[2]{T_{\COMM,#1,#2}^{(q)}}

\def\LAP{\mathcal{L}}
\newcommand{\ex}[1] {{\mathbb{E}}\left\{{#1}\right\}}

\newcommand{\ilaptt}[2]{\LAP_{#1}^{-1}\left[#2\right]}
\newcommand{\ilapt}[1]{\LAP_{#1}^{-1}\left(s\right)}
\newcommand{\fgbb}[1]{F_{\Gamma}\left(#1\right)} 
\newcommand{\zfx}[3]{#1^{2/#2}\int_{\left(#3/#1\right)^{2/#2}}^{\infty}\frac{1}{1+u^{#2/2}}du}

\newcommand{\hgf}[3]{{}_{#1}F_{#2}\!\left(#3\right)}

\newcommand{\sinrwik}[2]{\text{SIR}_{#1,#2}^{(q)}}

\newcommand{\epwik}[2]{\epsilon_{#1,#2}^{(q)}}
\newcommand{\epuikb}[2]{\overline{\epsilon}_{#1,#2}^{(u)}}
\newcommand{\epdikb}[2]{\overline{\epsilon}_{#1,#2}^{(d)}}
\newcommand{\epwikb}[2]{\overline{\epsilon}_{#1,#2}^{(q)}}
\newcommand{\epdikbh}[2]{\hat{\overline{\epsilon}}_{#1,#2}^{(d)}}

\newcommand{\bd}{\begin{description}}
\newcommand{\ed}{\end{description}}
\newcommand{\be}{\begin{enumerate}}
\newcommand{\ee}{\end{enumerate}}
\newcommand{\bi}{\begin{itemize}}
\newcommand{\ei}{\end{itemize}}
\newcommand{\bl}{\begin{list}}
\newcommand{\el}{\end{list}}
\newcommand{\bt}{\begin{tabbing}}
\newcommand{\et}{\end{tabbing}}

\newcounter{eqncnt}

\newcounter{eqnback}

\newcommand{\blue}[1]{{#1}}

\IEEEoverridecommandlockouts

\begin{document}

 

\title{Mobile Edge Computing-Enabled \\Heterogeneous Networks}

\author{
	\vspace{0.2cm}
	Chanwon~Park and
	Jemin~Lee, \textit{Member, IEEE}
	
	%
	%
	%
	\thanks{
		The material in this paper was presented, in part, at the International Conference on Communications, Kansas City, MO, May 2018 \cite{parlee:18}.
	} 
	\thanks{
		C.\ Park and J.\ Lee are with the    
		Department of Information and Communication Engineering,
		Daegu Gyeongbuk Institute of Science and Technology,
		Daegu, South Korea, 42988      
		(e-mail:\texttt{pcw0311@dgist.ac.kr}, \texttt{jmnlee@dgist.ac.kr}). 
	}
}

%

\maketitle 

%

%

%
%
\acresetall
\begin{abstract}
%
The \ac{MEC} has been introduced for providing computing capabilities at the edge of networks to improve the latency performance of wireless networks. In this paper, we provide the novel framework for \ac{MEC}-enabled \acp{HetNet}, composed of the multi-tier networks with \acp{AP} (i.e., \ac{MEC} servers), which have different transmission power and different computing capabilities. 
%
In this framework, we also consider multiple-type mobile users with different sizes of computation tasks, and they offload the tasks to a \ac{MEC} server, and receive the computation resulting data from the server. 
\blue{We derive the \ac{SECP}, defined as the probability that a user offloads and finishes its computation task at the \ac{MEC} server within the target latency.
We provide a closed-form expression of the approximated \ac{SECP} for general case, and closed-form expressions of the exact \ac{SECP} for special cases.
%
This paper then provides the design insights for the optimal configuration of \ac{MEC}-enabled \acp{HetNet} by analyzing the effects of network parameters and bias factors, used in \ac{MEC} server association, on the \ac{SECP}.
Specifically, it shows how the optimal bias factors in terms of \ac{SECP} can be changed according to the numbers of user types and tiers of \ac{MEC} servers, and how they are different to the conventional ones that did not consider the computing capabilities and task sizes.}
%
%
\end{abstract}


\acresetall


\section{Introduction} \label{sec:intro}

As a wireless communication is getting improved, mobile users are processing a numerous and complex computation tasks. To support the mobile users, the mobile cloud computing has been considered, which enables the centralization of the computing resources in the clouds. On the other hands, in recent years, the computation and battery capabilities of mobile users have been improved, which enables the mobile users to process the complex tasks. For that reason, the computation tasks start to be performed in the network edge including mobile users or servers located in small-cell \ac{AP} and it is called the \ac{MEC} \cite{macbec:17}.

One of the requirements of future wireless communications is the ultra-low latency. 
The cloud-radio access network (C-RAN) has been introduced to lower the computation latency of mobile users 
by making them offload complex tasks to a centralized cloud server \cite{trahajpanpom:17}.
However, to utilize the C-RAN, we need to experience inevitable long communication latency to reach to the far located central server. 
When the \ac{MEC} is applied, 
mobile users can compute the large tasks by offloading to the nearby \ac{MEC} servers, instead of the central server \cite{maoyouzhahualet:17}.
Although the computing capabilities of \ac{MEC} servers can be lower than those of the C-RAN servers, 
offloading tasks to \ac{MEC} servers can be more benefitial for some latency-critical applications
such as autonomous vehicles and sensor networks for health-care services.
Hence, the \ac{MEC} becomes one of the key technologies for future wireless networks. 

The performance of \ac{MEC} in wireless networks has been studied, mostly focusing on the minimization of energy consumption or communication and computing latency. 
Specifically, the energy minimization problem has been considered for proposing the policy of offloading to MEC servers with guaranteeing a certain level of latency \cite{youhuachakim:17, taootadonqili:17,dintanlaque:17, maozhasonlet:16, sarscubar:15, wanxuwancui:18, zhahuninngazhoweichehu:18}. 
Users with different computing capabilities are considered in \cite{youhuachakim:17},
and a single \cite{youhuachakim:17, taootadonqili:17} or multiple MEC servers \cite{dintanlaque:17} are used for each user.
\blue{The energy minimization problem has also been investigated for a energy harvesting user \cite{maozhasonlet:16}, multicell MIMO systems \cite{sarscubar:15}, and wireless-powered MEC server \cite{wanxuwancui:18}.
The tradeoff between energy consumption and latency is investigated by considering the residual energy of mobile devices and the energy-aware offloading scheme is proposed in \cite{zhahuninngazhoweichehu:18}.}

The latency minimization problem has been considered by analyzing the computation latency at \ac{MEC} servers using queueing theory \cite{liumaozhalet:16,zhazhoguoniu:17,kohanhua:17,chehao:18}.
The optimal offloading policy was presented for minimizing the mean latency \cite{liumaozhalet:16} or maximizing the probability of guaranteeing the latency requirements \cite{zhazhoguoniu:17}. The tradeoff between the latency and communication performance (i.e., network coverage) has also been presented in \cite{kohanhua:17}.
\blue{The task offloading problem for software-defined ultra-dense network is considered to minimize the average task duration with limited energy consumption of user in \cite{chehao:18}.
Recently, joint latency and energy minimization problem has also been studied by using the utility function defined based on both the energy and latency components \cite{zhaligonzha:19, lialiulokhua:19}.
In \cite{zhaligonzha:19}, a collaborative offloading problem is formulated to maximize the system utility by jointly optimizing the offloading decision and the computing resource assignment of \ac{MEC} servers for vehicular networks where \ac{MEC} and cloud computing are available.
In \cite{lialiulokhua:19}, the joint radio and computation resource allocation problem is considered to maximize the sum offloading rate and minimize the mobile energy consumption.}

However, except for \cite{zhazhoguoniu:17}, most of the prior works are based on the mean (or constant) computation latency, which fails to show the impact of latency distribution on the \ac{MEC} network performance. 
Furthermore, there is no work that considers the heterogeneous \ac{MEC} servers, which have different computing capabilities and transmission power, and various sizes of user tasks, impeding the efficient design of \ac{MEC}-enabled \ac{HetNet}. In the future, the \ac{MEC} will be applied not only to \acp{AP} or base stations (BSs) but to all computing devices around us such as mobile devices. Therefore, it is required to investigate how to design the \ac{MEC}-enabled network that has various types of \ac{MEC} servers, which is the main objective of this paper.

The \ac{HetNet} has been studied when \acp{AP} have different resources such as transmission power \cite{dhiganbacand:12,sindhiand:13,novdhiand:13,sinzhaand:15,josanxiaand:12, zhayanquelee:17}, mainly by focusing on the communication performance, not the computing performance.
In most of the works, the stochastic geometry has been applied for the spatial model of distributed users and \acp{AP} using \acp{PPP} \cite{BacBla:09a}.
For example, the baseline model containing the outage probability and average rate for downlink is shown in \cite{dhiganbacand:12}. The network modeling and coverage analysis are provided in \cite{sindhiand:13} for downlink, in \cite{novdhiand:13} for uplink, and in \cite{sinzhaand:15} for decoupling of uplink and downlink. 
%
%
The cell range expansion for load balancing among \acp{AP} is considered in \cite{sindhiand:13} and \cite{josanxiaand:12}. 
The \acp{HetNet} with line-of-sight and non-line-of-sight link propagations are also investigated in \cite{zhayanquelee:17} and \cite{choliuleenohque:18}.
Recently, the stochastic geometry has also been applied for the performance analysis of randomly distributed \ac{MEC} servers in \cite{kohanhua:17}, but the latency distribution, heterogeneous \ac{MEC} servers, and various sizes of user tasks are not considered for the design of \ac{MEC}-enabled networks. 

In this work, we provide the novel framework of the \ac{MEC}-enabled \acp{HetNet}.
We consider the multi-tier networks composed of \ac{MEC} servers having different computing capacities and the multi-type users having different computation task sizes.
\blue{To evaluate the network performance, the \ac{SECP} is defined as the probability that a user offloads and finishes its computation task at the \ac{MEC} server within the target latency.
We provide the \ac{SECP} for general cases by applying the queueing theory to analyze the computation latency distribution and the stochastic geometry to analyze the both uplink and downlink communication latency.}
The main contribution of this paper can be summarized as below.

\bi
\item We develop the novel framework of the \ac{MEC}-enabled \acp{HetNet} characterized by the \emph{multi-tier} \ac{MEC} server having different computing capacities and \emph{multi-type} users having different computation task sizes.
\item \blue{We introduce and derive the SECP of the \ac{MEC}-enabled \acp{HetNet}, which considers both the computing and communication performance. To the best of our knowledge, this is the first performance analysis of the MEC-enabled networks, which consider the \emph{multiple type} users. We also provide the closed-form expression of the approximated SECP for general cases, and the closed-form expression of the exact SECP for special cases.}
\item \blue{We analyze how the association bias factors for each \ac{MEC} tiers and each user types affect the SECP, and provide design insights on the bias factors for different network parameters including the numbers of user types and tiers of \ac{MEC} servers.}

%
\ei

\begin{figure}[t!]
	\begin{center}   
		{ 
			\includegraphics[width=1.00\columnwidth]{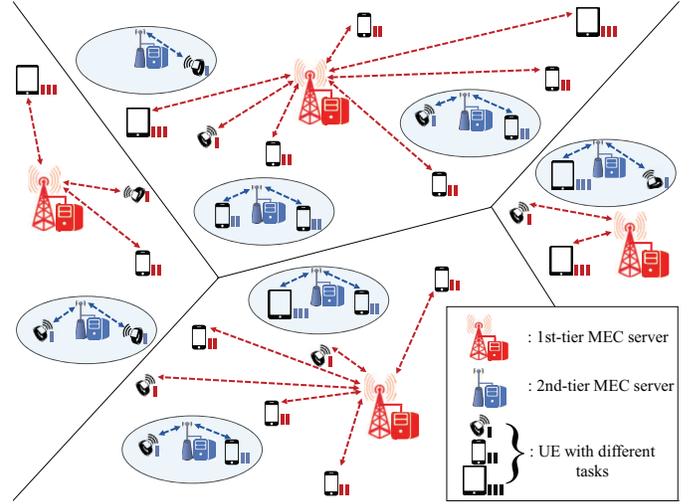}
		}
	\end{center}
	\vspace{-2mm}
	\caption{
		The \ac{MEC}-enabled \ac{HetNet} composed of the multi-tier \ac{MEC} servers and the multi-type users with different size of tasks.
	}
	\vspace{0mm}
	\label{fig:network}
\end{figure}

The remainder of this paper is organized as follows. Section \ref{sec:models} describes the \ac{MEC}-enabled \ac{HetNet} model and gives the queueing model for calculating the latency. \blue{Section \ref{sec:pcm} analyzes the communication latency. Section \ref{sec:ps} defines and analyzes the \ac{SECP}. Section \ref{sec:numresult} presents the effects of association bias and network parameters on \ac{SECP}.} Finally, conclusions are given in Section \ref{sec:conclusions}.


\section{MEC-enabled Heterogeneous Network Model} \label{sec:models}


In this section, we present the network model and the latency model of the \ac{MEC}-enabled \ac{HetNet}.

\subsection{Network Model}

We consider a \ac{MEC}-enabled \ac{HetNet}, which is composed of $\mathbf{K}$ tiers of \ac{MEC} servers, located in \acp{AP}. An example of the network is given in Fig. \ref{fig:network}.
\ac{MEC} servers in different tiers have different transmission and computing capabilities. 
We use $\SetK=\{1,\cdots,\mathbf{K}\}$ as the index set of $\mathbf{K}$ tiers of \ac{MEC} servers.
%
%
The \ac{MEC} servers are distributed according to a homogeneous \ac{PPP} $\pppm$ with spatial density $\lbdm$. The locations of the $k$-tier \ac{MEC} servers are also modeled as a homogeneous \ac{PPP} $\pppmk{k}$ with spatial density $\lbdmk{k}=\pmk{k}\lbdm$ 
where $\pmk{k}$ is the portion of the $k$-tier \ac{MEC} servers. 
Each servers are assumed to have one CPU and one queue, which has an infinite waiting space. 
\ac{MEC} server in a $k$-tier transmits with the power $\pwrmk{k}$. 
A channel is assigned to one user only in the cell of \ac{MEC} server (AP), 
and the ratio of users using the same uplink channel is $\kappa$, which denotes the frequency reuse factor. The density of uplink interfering users offloading to a $k$-tier \ac{MEC} server is then given by $\kappa\lbdmk{k}$.


Computing the large computation task at a mobile device may not be finished within a required time. Hence, users offload their tasks to the \ac{MEC} servers, which compute/process the tasks and send the resulting data back to the user.
Users are categorized into $\mathbf{I}$ types according to the size of their offloading tasks. 
The $\SetI=\{1,2, \cdots,\mathbf{I}\}$ denotes the index set of $\mathbf{I}$ type users. 
\blue{A $i$-type user offloads the computation task with $\dit{i}{(c)}$ packets to a \ac{MEC} server.
The packet in computation tasks consists of $\punit$ bits.
We assume that the both the request message and the resulting data are proportional to the computation task size for simplicity.
For that, the user transmits the computation request message in $\dit{i}{(r)}$ packets to the \ac{MEC} server with the power $\pwrui{i}$ (i.e., uplink transmission), and receives the computation resulting data in $\dit{i}{(d)}$ packets (i.e., downlink transmission).}

Each transmission is happened in a time slot $\ts$.\footnote{Note that the whole uplink and downlink transmission time can consist of different number of time slots. Here, we focus on the transmission in one time slot $\ts$ for both uplink and downlink, and the arrival task rate at a \ac{MEC} server is defined as the number of arriving requests per $\ts$.}
The locations of users are modeled as a homogeneous \ac{PPP} $\pppu$ with spatial density $\lbdu$. The $i$-type users are distributed according to a homogeneous \ac{PPP} $\pppui{i}$ with spatial density $\lbdui{i}=\pui{i}\lbdu$, where $\pui{i}$ is the portion of the $i$-type users. 

\blue{Besides the uplink and downlink transmission time, the computation latency at the \ac{MEC} server is caused when the computation tasks are offloaded to a MEC server. Hence, when an $i$-type user offloads to a $k$-tier \ac{MEC} server, the total latency of an $i$-type user can be defined as 
\begin{align} \label{eq:ttotal}
\tti{i}=
\tcmuik{i}{k}+\tcmdik{i}{k}+\tcik{i}{k}
\end{align}
%
for $k\geq1$ where $\tcik{i}{k}$ is the computation latency of an $i$-type user at a $k$-tier \ac{MEC} server. In \eqref{eq:ttotal}, $\tcmuik{i}{k}$ and $\tcmdik{i}{k}$ are the expected uplink transmission time and downlink transmission time, respectively.} In this paper, when an $i$-type user offload to a $k$-tier \ac{MEC} server, $\tcik{i}{k}$ is given by $\tcik{i}{k} = \tquk{k}+\tsvik{i}{k}$, where $\tquk{k}$ is the waiting time at the queue and $\tsvik{i}{k}$ is the processing (service) time. To analyze the total latency, both communication latency (i.e., transmission time) and computation latency need to be considered.

%


%
\begin{figure}[t!]
	\begin{center}   
		{ 
			\includegraphics[width=1.00\columnwidth]{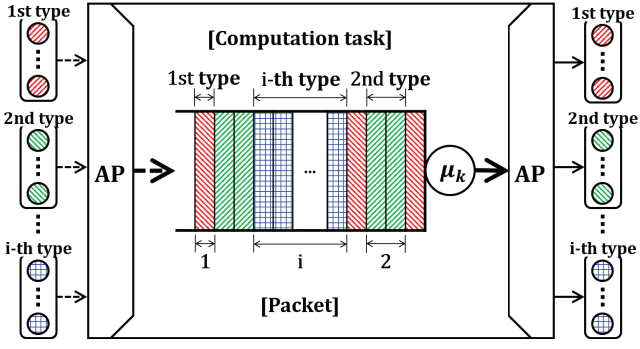}
		}
	\end{center}
		 \vspace{-2mm}
	\caption{
		A $k$-tier \ac{MEC} server when the users offload their computation tasks and receive their resulting data.
	}
		 \vspace{0mm}
	\label{fig:queue}
\end{figure}
%


\begin{table}
	\caption{Notations used throughout the paper.} \label{table:notation}
	\vspace{-7mm}
	\begin{center}
		\rowcolors{2}
		{cyan!15!}{}
		\renewcommand{\arraystretch}{1.3}
		\begin{tabular}{ c  p{7cm} }
			\hline 
			{\bf Notation} & {\hspace{2.5cm}}{\bf Definition}
			\\
			\midrule
			\hline
			$\betawik{i}{k}$	& Rate coverage probability threshold that an $i$-type user offloads to a $k$-tier \ac{MEC} server for $\text{u, d}$ (u=uplink, d=downlink)\\ \addlinespace
			$\biasik{i}{k}$	& Bias factor of an $i$-type user offloading to a $k$-tier \ac{MEC} server\\ \addlinespace
			$\dit{i}{(c)}$ 		& Computation task size of an $i$-type user \\ \addlinespace
			$\epwik{i}{k}$ 	& Target data rate of an $i$-type user offloading to the $k$-tier \ac{MEC} server for $q=\{\text{u, d}\}$ (u=uplink, d=downlink)\\ \addlinespace
			$\lbdmk{k}$ 	& Spatial density of a $k$-tier \ac{MEC} server \\ \addlinespace
			$\lbdui{i}$ 	& Spatial density of an $i$-type user \\ \addlinespace
			$\muk{k}$		& Service rate of a $k$-tier \ac{MEC} server \\ \addlinespace 
			$\lbdaik{i}{k}$		& Arrival rate of an $i$-type user offloading to a $k$-tier \ac{MEC} server \\ \addlinespace 
			$\pwrui{i}$		& Transmission power of an $i$-type user\\ \addlinespace
			$\pwrmk{k}$ 	& Transmission power of an $k$-tier \ac{MEC} server\\ \addlinespace
			$\pui{i}$ 		& Portion of an $i$-type users \\ \addlinespace 
			$\pmk{k}$ 		& Portion of a $k$-tier \ac{MEC} servers \\ \addlinespace
			$\poik{i}{k}$		& Probability that an $i$-type user offloads to a $k$-tier \ac{MEC} server\\ \addlinespace
			$\pecik{i}{k}$		& Successful edge computing probability that an $i$-type user offloads to a $k$-tier \ac{MEC} server\\ \addlinespace
			%
			$\pppmk{k}$ & PPP for a $k$-tier \ac{MEC} server distribution \\ \addlinespace
			$\pppui{i}$ & PPP for an $i$-type user distribution \\ \addlinespace
			%
			$\tcmwik{i}{k}$	& Communication latency of an $i$-type user offloading to a $k$-tier \ac{MEC} server for $q=\{\text{u, d}\}$ (u=uplink, d=downlink)\\ \addlinespace
			$\tsvik{i}{k}$	& Service time of an $i$-type user offloading to a $k$-tier \ac{MEC} server \\ \addlinespace
			$\ttgi{i}$ & Target latency of an $i$-type user\\ \addlinespace 
			$\tquk{k}$		& Waiting time at a $k$-tier \ac{MEC} server\\ \addlinespace
			$\weightik{i}{k}$	& Weighting factor of a $i$-type user offloading to a $k$-tier \ac{MEC} server\\ \addlinespace 
			$\BW^{(q)}$		& Bandwidth for $q=\{\text{u, d}\}$ (u=uplink, d=downlink) transmission \\ 
			\hline 
		\end{tabular}
	\end{center}\vspace{-2mm}
\end{table}%

\subsection{\blue{Communication Latency Model}}

\blue{
In a \ac{MEC}-enabled \ac{HetNet}, offloading users have communication latency, when the computation request message are transmitted through uplink channel and its resulting data are received through downlink channel. When an $i$-type user offloads to a $k$-tier \ac{MEC} server, the rate coverage probability is defined as the probability that the acheivable data rate $\drwik{i}{k}$ is greater than and equal to the target data rate $\epwik{i}{k}$ given by
\begin{align} \label{eq:reli}
\pcmqik{i}{k}=\px{\drwik{i}{k}\geq\epwik{i}{k}}
\end{align}
for $q=\{\text{u}, \text{d}\}$ where u and d indicate the uplink and downlink channels, respectively. If the data rate of the user is less than the target data rate, the data transmission becomes failed. We define the maximum target data rate $\epwikb{i}{k}$ guaranteeing a certain level of coverage probability as \eqref{eq:epwikb}
\begin{align} \label{eq:epwikb}
\pcmqik{i}{k}=\px{\drwik{i}{k}\geq\epwikb{i}{k}}\geq\betawik{i}{k}
\end{align}
where $\betawik{i}{k}$ is the target coverage probability. Here the retransmission from communication failure is not considered, which can also be ignored by setting high $\betawik{i}{k}$ (i.e., ignorable communication failures)\footnote{\blue{If the data retransmission is considered, the temporal and spatial correlation of interference received during retransmission exist, which makes the communication latency non-tractable \cite{zhoquege:17}, \cite{yanque:18}. In addition, interacting queueing status of transmitting users need to be considered in the computation latency analysis, which also further complicate the analysis. Hence, as analyzing the effect of retransmission in \ac{MEC}-enabled network is beyond the scope of our work, the retransmission is neglected in this work}} From \eqref{eq:epwikb}, we then define the communication latency of $i$-type user for offloading to the $k$-tier \ac{MEC} server, denoted by $\tcmwik{i}{k}$, as
\begin{align} \label{eq:tcmwik}
\tcmwik{i}{k}=\frac{\dit{i}{(r)}\punit}{\epwikb{i}{k}}.
\end{align}
}
 
\subsection{\blue{Computation Latency Model}}

\blue{Besides the communication latency, offloading users in \ac{MEC}-enabled \ac{HetNet} also have the computation latency caused when the computation tasks are computed, i.e., processed in the \ac{MEC} server in \ac{AP}.} To analyze the computation latency, the arrival rate of the computation tasks and the service time distribution at the \ac{MEC} server need to be determined.

Since the users are distributed as a \ac{PPP}, the arrivals of computing tasks at the server follow a Poisson process with a certain arrival rate. Here, the arrival rate for an $i$-type user task to a $k$-tier \ac{MEC} server, denoted by $\lbdaik{i}{k}$, is determined as
\begin{align}\label{eq:arvrate}
\blue{\lbdaik{i}{k}=\frac{\betauik{i}{k}\lbdu\pui{i}\poik{i}{k}}{\lbdm\pmk{k}}} 
\end{align}
where $\poik{i}{k}$ is the probability that an $i$-type user offloads to a $k$-tier \ac{MEC} server. When a user offload the computation tasks, we assume a user select a \ac{MEC} server using the association rule, which is based on the biased average receive power, defined as \cite{sindhiand:13}
\begin{align} \label{eq:association}
k = \arg \max_{j\in \SetK} 
\left\{ 
\max_{\x_j  \in \pppmk{j}}\weightik{i}{j} \zik{\lx_\text{o}}{\x_j}^{-\alpha}
\right\}
\end{align}
where $\zik{\lx}{\ly}$ is the distance between $\lx$ and $\ly$, $\alpha$ is the pathloss exponent,
$\weightik{i}{j}=\pwrmk{j}\biasik{i}{j}$ is the weighting factor for an $i$-type user offloading to a $j$-tier \ac{MEC} server, and $\biasik{i}{j}$ is the bias factor. 
This shows that the $i$-type user located at $\lx_{\text{o}}$ is associated to the $k$-tier \ac{MEC} server. 
Based on the association rule in \eqref{eq:association}, $\poik{i}{k}$ in \eqref{eq:arvrate} is given by \cite{sindhiand:13}
%
\begin{align} \label{eq:poik}
\poik{i}{k}
= 2\pi\lbdmk{k}\int_{0}^{\infty}x\exp\left\{-\pi x^{2}\sum_{j \in \SetK}\lbdmk{j}\weightikh{i}{j}^{2/\alpha}\right\}dx
\end{align}
%
where $\weightikh{i}{j}$ is $\weightik{i}{j}/\weightik{i}{k}$. 

\blue{The service rate (i.e., computing capability) of a $k$-tier \ac{MEC} server $\muk{k}$ is determined as $\muk{k}=F_{\text{m},k}/\left(C_{\text{u}}\punit\right)$ where $C_{\text{u}}$ is the number of CPU cycles required for computing $1$ bit of a computation task and $F_{\text{m},k}$ is the computing capacity of $k$-tier \ac{MEC} server measured by the number of CPU cycles per second.}
The distribution of the service time for one packet in a $k$-tier \ac{MEC} server is modeled as the exponential distribution with $1/\muk{k}$. Since an $i$-type user offloads $\dit{i}{(c)}$ packets of the task, $\tsvik{i}{k}$ follows the Erlang distribution, and the \ac{pdf} of $\tsvik{i}{k}$, $\fsvik{i}{k}$, is given by
\begin{align} \label{eq:svtikpdf}
\fsvik{i}{k}
=\frac{\muk{k}^{\dit{i}{(c)}}t^{\dit{i}{(c)}-1}\exp\left(-\muk{k} t\right)}{\left(\dit{i}{(c)}-1\right)!}.
\end{align}
The service time in a $k$-tier \ac{MEC} server, denoted by $\tsvk{k}$, is the weighted sum of $\tsvik{i}{k}$ given by $\tsvk{k}=\sum_{i\in\SetI}\left(\lbdaik{i}{k}/\lbdak{k}\right)\tsvik{i}{k}$ where $\lbdak{k}=\sum_{i\in\SetI}\lbdaik{i}{k}$ is the arrival rate of users offloading to a $k$-tier \ac{MEC} server. 
Hence, the \ac{pdf} of $\tsvk{k}$ is given by
\begin{align}	
\fsvk{k}=\sum_{i\in\SetI}\left(\frac{\lbdaik{i}{k}}{\lbdak{k}}\right)\fsvik{i}{k}.
\end{align}

\subsection{\blue{Performance Metric}}

\blue{
For \ac{MEC}-enabled \acp{HetNet}, we derive the \ac{SECP} as the performance metric. The \ac{SECP} is the probability that the computation in \ac{MEC} server and communication between \ac{MEC} server and mobile users are finished within a target latency. According to \eqref{eq:ttotal}, the \ac{SECP} for an $i$-type user offloading to a $k$-tier \ac{MEC} server, denoted by $\pecik{i}{k}$, is defined as
\begin{align} \label{eq:pecik}
\pecik{i}{k}
=\px{\tquk{k}+\tsvik{i}{k}+\tcmuik{i}{k}+\tcmdik{i}{k}\leqq\ttgi{i}}
\end{align}
where $\ttgi{i}$ is the target latency of a $i$-type user. The \ac{SECP} can be considered equivalently to 1) the probability that a user can finish the edge computation within the target latency, and 2) the average fraction of mobile users that satisfy the latency requirements, i.e., latency QoS. Using the law of total probability, the overall \ac{SECP} for the network is given by
\begin{align} \label{eq:ps}
\pec=\sum_{k\in\SetK}\sum_{i\in\SetI}\pui{i}\poik{i}{k}\pecik{i}{k}.
\end{align}
}

\section{\blue{Communication Latency Analysis}} \label{sec:pcm}

\blue{
In this section, we analyze the communication latency i.e., uplink and downlink transmission time, when an $i$-type user communicates with a $k$-tier \ac{MEC} server. We assume that the channel state information for both users and \ac{MEC} servers are known, so that both users and \ac{MEC} servers can determine the data rate satisfying the target coverage probability before transmiting the computation tasks or results. According to \eqref{eq:tcmwik}, the maximum target data rate needs to be determined first to obtain the both uplink and downlink transmission time. The maximum target data rate $\epwikb{i}{k}$ is given by
\begin{align} \label{eq:defepwikb}
\epwikb{i}{k}=\int_{y>0}\epwikb{i}{k}\left(y\right)f_{Y_{i,k}}\left(y\right)dy
\end{align}
for $q=\{\text{u, d}\}$ where $f_{Y_{i,k}}\left(y\right)$ is the \ac{pdf} of the distance between the $i$-type user and the associated $k$-tier \ac{MEC} server denoted by $Y_{i,k}$, given by \cite[Lemma 4]{sindhiand:13}
\begin{align} \label{eq:fyik}
f_{Y_{i,k}}\left(y\right)
=\frac{2\pi\lbdmk{k}}{\poik{i}{k}}
y\exp
\left\{
-\pi\sum_{j\in\SetK}
\lbdmk{j}\weightikh{i}{j}^{2/\alpha}
y^{2}
\right\}.
\end{align}
In \eqref{eq:defepwikb}, $\epwikb{i}{k}\left(y\right)$ is the maximum target data rate when an $i$-type user offloads to and receives from a $k$-tier \ac{MEC} server located at $y$. The $\epwikb{i}{k}\left(y\right)$ is given by guaranteeing the rate coverage probability $\pcmqik{i}{k}$ in \eqref{eq:epwikb}. The $\pcmqik{i}{k}$ can be presented by 
\begin{align} \label{eq:scpsir}
\pcmqik{i}{k}=\px{\sinrwik{i}{k}\left(y\right)\geq 2^{\frac{\epwikb{i}{k}}{\BW^{(u)}}}\!\!-1}\geq\betawik{i}{k}
\end{align}
for $q=\{\text{u, d}\}$ where $\!\BW^{(q)}\!$ is the bandwidth, and $\sinrwik{i}{k}$ is the received \ac{SIR} when the $i$-type user offloads to and receives from the $k$-tier \ac{MEC} server located at $y$.} For the analytical tractability, some assumptions are used here to derive the uplink transmission time.

1) \emph{Assumption 1}: The distribution of uplink interfering users follows the \ac{PPP}.

The uplink interfering user distribution is not a \ac{PPP} because locations of the users using same uplink channel is from the dependent thinning of the \ac{AP} locations. However, according to \cite{novdhiand:13}, such effect can be weak. Hence, we use this assumption, as in other papers \cite{novdhiand:13, elshos:14}, for analysis tractability.

2) \emph{Assumption 2:} Uplink and downlink interference are independent. 

Uplink and downlink interference are not independent because the locations of \ac{MEC} servers and interfering users are dependent. Although some papers like \cite{leeque:15} considered this dependence by a simplified method, it is still complicate to analyze the dependence. As the uplink analysis is not main objective of this work, we apply this assumption.\footnote{There is a recent results in \cite{monwasdas:17}, which provides the uplink performance with less assumptions \blue{by characterizing the distribution of active uplink users. However, the results still need a lot of mathematical operations, which makes the results hard to maintain the analytical tractability. Hence,} we note that applying more accurate analysis does not change our framework.}

\blue{
Using those assumptions and Theorem 1 in \cite{novdhiand:13}, the uplink rate coverage probability (i.e., $q=u$) is given by
%
%
\begin{align} \label{eq:sinriku1}
\pcmuik{i}{k}=
\LAP_{I_{i,k}^{(u)}}\left\{y^{\alpha}\pwrui{i}^{-1}\left(2^{\frac{\epuikb{i}{k}}{\BW^{(u)}}}-1\right)\right\}
\end{align}
where $I_{i,k}^{(u)}$ is the interference when an $i$-type user offloads to a $k$-tier \ac{MEC} server. 
The $\LAP_{I_{i,k}^{(u)}}\left(s\right)$ in \eqref{eq:sinriku1} is presented as \cite{josanxiaand:12}
%
\begin{align} \label{eq:sinriku2}
\LAP_{I_{i,k}^{(u)}}\left(s\right)
=
\exp\left\{\!-2\pi\lbdm\int_{z_{i,k}}^{\infty}\frac{x}{1+\left(s\pwrui{i}\right)^{-1}x^{\alpha}}dx\right\}
\end{align}
where $z_{i,k}$ is the distance to the nearest $k$-tier \ac{MEC} server unassociated with the $i$-type user. Since the nearest interfering user can be closer than the associated user, $z_{i,k}$ becomes zero. By replacing $s=y^{\alpha}\pwrui{i}^{-1}\left(2^{\epuikb{i}{k}/\BW^{(u)}}\!\!-1\right)$ in \eqref{eq:sinriku2} and $z_{i,k}=0$, \eqref{eq:sinriku1} is calculated by
\begin{align} \label{eq:sinriku3}
\pcmuik{i}{k}
\!\!=
&\exp\!\left\{\!\!-2\pi\kappa\lbdm\!\!\int_{0}^{\infty}\!\!\!\!\!\!\frac{x}{1+\left(y^{-\alpha}/\!\left(2^{\epuikb{i}{k}/\BW^{(u)}}\!\!-1\right)\!\right)\!x^{\alpha}}dx\!\right\}\nonumber \\
=&\exp{\left\{-\pi\kappa\lbdm y^{2}Z\left(2^{\epuikb{i}{k}/\BW^{(u)}}\!\!-1,\alpha,0\right)\right\}}
\end{align}
where $Z\left(a,b,c\right)=\zfx{a}{b}{c}$.
}
\blue{
It is hard to calculate $\pcmuik{i}{k}$ for the general path loss. However, for the path loss factor $\alpha=4$, $\pcmuik{i}{k}$ can be presented in a tractable form. Using \eqref{eq:defepwikb} and \eqref{eq:sinriku3}, the uplink maximum target data rate is derived in the following lemma.
\begin{lemma} \label{lma:maxtgrate_up}
For the uplink transmission with $\alpha=4$, the maximum target data rate $\epuikb{i}{k}$ is given by
\begin{align} \label{eq:comepuikb}
\epuikb{i}{k}=&
\frac{\pi\lbdmk{k}\BW^{(u)}}{\poik{i}{k}\ln 2}
\biggr[
\frac{2}{\delta_{1}^{(u)}}
\left\{
\ln\left(\delta_{2}^{(u)}\right)
-\text{ci}\left(\delta_{2}^{(u)}\right)
\cos\left(\delta_{2}^{(u)}\right)\right. \nonumber \\
&
\left.
-\text{si}\left(\delta_{2}^{(u)}\right)
\sin\left(\delta_{2}^{(u)}\right)
+\mathbf{C}
\right\}
\biggr]
\end{align}
where $\delta_{1}^{(u)}$ and $\delta_{2}^{(u)}$ are given, respectively, by
%
%
\begin{align}
\delta_{1}^{(u)}=\pi\sum_{j\in\SetK}\lbdmk{j}\weightikh{i}{j}^{1/2},\indent
\delta_{2}^{(u)}=\frac{-2\delta_{1}^{(u)}\ln\betauik{i}{k}}{\pi^{2}\kappa\lbdmk{k}}.
\end{align}
In \eqref{eq:comepuikb}, $\text{ci}\left(\cdot\right)$, $\text{si}\left(\cdot\right)$, and $\mathbf{C}$ are cosine integral function, sine integral function, and euler constant, respectively.
\begin{IEEEproof}
For $\alpha=4$, $\pcmuik{i}{k}$ in \eqref{eq:sinriku3} can be presented as
\begin{align} \label{eq:pcmuika4}
\pcmuik{i}{k}
=\exp{\left\{-\frac{\pi^{2}}{2}\kappa\lbdm y^{2}\left(2^{\frac{\epuikb{i}{k}}{\BW^{(u)}}}\!\!-1\right)^{1/2}\right\}}.
\end{align}
From \eqref{eq:pcmuika4} and \eqref{eq:scpsir}, $\epuikb{i}{k}\left(y\right)$ is given by
\begin{align} \label{eq:epuikby}
\epuikb{i}{k}\left(y\right)=\BW^{(u)}\log_{2}\left\{1+\left(\frac{-2\ln\betauik{i}{k}}{\pi^{2}\kappa\lbdmk{k}y^{2}}\right)^{2}\right\}.
\end{align}
Substituting \eqref{eq:epuikby} and \eqref{eq:fyik} into \eqref{eq:defepwikb} and replacing $y$ in $t$ according to $t=y^{2}$, $\epuikb{i}{k}$ is presented as
\begin{align} \label{eq:comepuikbprior}
\epuikb{i}{k}
=& 
\frac{\pi\lbdmk{k}\BW^{(u)}}{\poik{i}{k}\ln 2}
\!\!\int_{0}^{\infty}
\!\!\left[
\ln\left\{\!t^{2}\!+\!\left(\!\frac{-2\ln\betauik{i}{k}}{\pi^{2}\kappa\lbdmk{k}}\!\right)^{\!\!\!2}\right\}\!\!-\!\ln\left\{t^{2}\right\}\!
\right]
\nonumber \\
&\times \exp
\left\{
-\pi\sum_{j\in\SetK}
\lbdmk{j}\weightikh{i}{j}^{2/\alpha}
t
\right\}
dt.
\end{align}
From the equations in \cite[eq. (4.331), eq. (4.338)]{GraRyz:B07}, \eqref{eq:comepuikbprior} becomes \eqref{eq:comepuikb}.
\end{IEEEproof}
\end{lemma}
}
\blue{
From the analysis in \cite{sindhiand:13} and \cite{josanxiaand:12}, the downlink rate coverage probability (i.e., $q=d$) is given by
\begin{align} \label{eq:sinrikd1}
\pcmdik{i}{k}
=
\prod_{j\in\SetK}\LAP_{I_{i,j}^{(d)}}\left\{y^{\alpha}\pwrmk{k}^{-1}\left(2^{\frac{\epdikb{i}{k}}{\BW^{(d)}}}-1\right)\right\}
\end{align}
where $I_{i,j}^{(d)}$ is the interference when an $i$-type user receives from a $j$-tier \ac{MEC} server.
In \eqref{eq:sinrikd1}, $\LAP_{I_{i,j}^{(d)}}\left(s\right)$ is given by substituting $\pwrui{i}$ and $z_{i,k}$ in \eqref{eq:sinriku2} into $\pwrmk{j}$ and $\weightikh{i}{j}^{1/\alpha}y$, respectively. By replacing $s=y^{\alpha}\pwrmk{k}^{-1}\left(2^{\epdikb{i}{k}/\BW^{(d)}}\!\!-1\right)$, \eqref{eq:sinrikd1} is represented by
\begin{align} \label{eq:sinrikd3}
&\pcmdik{i}{k} \nonumber \\
&=
\exp\!\left\{\!\!-\!\!\sum_{j\in\SetK}\!2\pi\lbdmk{j}\!\!\!\!\right.
\left.\int_{\!z_{i,j}}^{\infty}\!\!\!\frac{x}{1\!+\!\!\left(\!\pwrmkh{j}^{-1}y^{-\alpha}\!/\!\!\left(2^{\epdikb{i}{k}/\BW^{(d)}}\!\!\!-\!1\!\right)\!\right)\!\!x^{\alpha}}dx\!\!\right\} \nonumber \\
&=\exp{
	\!\left\{\!-\!\!\sum_{j\in\SetK}\!\pi \pwrmkh{j}^{2/\alpha}\!\lbdmk{j}y^{2}
	Z\!
	\left(
	2^{\epdikb{i}{k}/\BW^{(d)}}\!\!\!\!-\!1,\alpha,\!\biasikh{i}{j}
	\right)
	\!\!\right\}}
\end{align}
where $\pwrmkh{j}$ is $\pwrmk{j}/\pwrmk{k}$ and $\biasikh{i}{j}$ is $\biasik{i}{j}/\biasik{i}{k}$.
}
\blue{
Similar with the uplink case, $\pcmdik{i}{k}$ can be obtained in a tractable form for the path loss factor $\alpha=4$. However, since $\epdikb{i}{k}\left(y\right)$ is included in the $Z$ function in \eqref{eq:sinrikd3}, it is difficult to present $\epdikb{i}{k}$. For the analytical tractability, we obtain the lower bound of $\epdikb{i}{k}$ by approximating $\biasikh{i}{j}$ as $0$. Using \eqref{eq:defepwikb} and \eqref{eq:sinrikd3}, the lower bound of downlink maximum target data rate is derived in the following lemma.
\begin{lemma}
For the downlink transmission with $\alpha=4$, the lower bound of the maximum target data rate $\epdikbh{i}{k}$ is given by
\begin{align} \label{eq:comepdikb}
\epdikbh{i}{k}=&
\frac{\pi\lbdmk{k}\BW^{(d)}}{\poik{i}{k}\ln 2}
\biggr[
\frac{2}{\delta_{1}^{(d)}}
\left\{
\ln\left(\delta_{2}^{(d)}\right)
-\text{ci}\left(\delta_{2}^{(d)}\right)
\cos\left(\delta_{2}^{(d)}\right)\right. \nonumber \\
&
\left.
-\text{si}\left(\delta_{2}^{(d)}\right)
\sin\left(\delta_{2}^{(d)}\right)
+\mathbf{C}
\right\}
\biggr]
\end{align}
where $\delta_{1}^{(d)}$ and $\delta_{2}^{(d)}$ are given, respectively, by
%
%
\begin{align}
\delta_{1}^{(d)}=\pi\!\sum_{j\in\SetK}\!\lbdmk{j}\weightikh{i}{j}^{1/2},\indent
\!\!\!\!\delta_{2}^{(d)}=\!\frac{-2\delta_{1}^{(d)}\ln\betadik{i}{k}}{\pi^{2}\sum_{j\in\SetK}\lbdmk{j}\pwrmkh{j}^{1/2}}.
\end{align}
In \eqref{eq:comepuikb}, $\text{ci}\left(\cdot\right)$, $\text{si}\left(\cdot\right)$, and $\mathbf{C}$ are cosine integral function, sine integral function, and euler constant, respectively.
\begin{IEEEproof}
For $\alpha=4$, $\pcmdik{i}{k}$ in \eqref{eq:sinrikd3} can be presented as
\begin{align} \label{eq:sinrikd3app}
\pcmdik{i}{k}
\!\geq\!
\exp{
	\left\{-\sum_{j\in\SetK}\frac{\pi^{2}}{2}\pwrmkh{j}^{1/2}\lbdmk{j}y^{2}\!\!
	\left(
	2^{\frac{\epdikb{i}{k}}{\BW^{(d)}}}\!\!-1
	\right)^{\!\!1/2}
	\right\}}.
\end{align}
From \eqref{eq:sinrikd3app} and \eqref{eq:scpsir}, $\epdikbh{i}{k}$ is obtained by
\begin{align} \label{eq:epdikby}
\epdikbh{i}{k}\left(y\right)\!=\!\BW^{(d)}\log_{2}\!\left[1+\!\left(\frac{-2\ln\betadik{i}{k}}{\pi^{2}y^{2}\sum_{j\in\SetK}\pwrmkh{j}^{1/2}\lbdmk{j}}\right)^{\!2}\right].
\end{align}
Substituting \eqref{eq:epdikby} and \eqref{eq:fyik} into \eqref{eq:defepwikb} and replacing $y$ in $t$ according to $t=y^{2}$, $\epdikbh{i}{k}$ is presented as
\begin{align} \label{eq:comepdikbhprior}
\epdikbh{i}{k}
=& 
\frac{\pi\lbdmk{k}\BW^{(d)}}{\poik{i}{k}\ln 2}
\!\!\int_{0}^{\infty}
\biggr[
\ln\left\{\!t^{2}\!+\!\left(\!\frac{-2\ln\betadik{i}{k}}{\pi^{2}\sum_{j\in\SetK}\lbdmk{j}\pwrmkh{j}^{1/2}}\!\right)^{\!\!\!2}\right\}\!\! \nonumber \\
&-\!\ln\left\{t^{2}\right\}\!
\biggr]
\exp
\left\{
-\pi\sum_{j\in\SetK}
\lbdmk{j}\weightikh{i}{j}^{2/\alpha}
t
\right\}
dt.
\end{align}
From the equations in \cite[eq. (4.331), eq. (4.338)]{GraRyz:B07}, \eqref{eq:comepdikbhprior} becomes \eqref{eq:comepdikb}.
\end{IEEEproof}
\end{lemma}
}
\blue{
By substituting \eqref{eq:comepuikb} and \eqref{eq:comepdikb} into \eqref{eq:tcmwik}, we can derive both uplink and downlink transmission time (i.e., communication latency). Here, we use the lower bound of the target data rate for downlink as it can also guarantee the target coverage probability. From \eqref{eq:tcmwik}, we can show that the communication latency can be changed according to the target coverage probability.
}

\section{Successful Edge Computing Probability Analysis} \label{sec:ps}

In this section, we analyze the \ac{SECP} for an $i$-type user offloading to a $k$-tier \ac{MEC} server, defined in section \ref{sec:models}. Using \eqref{eq:pecik} and \cite{kle:75}, $\pecik{i}{k}$ is derived in the following theorem.
\begin{theorem} \label{trm:pcompik}
The \ac{SECP} for an $i$-type user offloading to a $k$-tier \ac{MEC} server, denoted by $\pecik{i}{k}$, is given by
%
\begin{align} \label{eq:pcompikint}
\pecik{i}{k} =&\int_{0}^{\infty}\!\!\!\int_{0}^{\blue{T_{\text{th},i}-r}}
\!\!\!\ilaptt{\tquk{k}}{
	\frac{\left(1-\rhok{k}\right)s}
	{ s-\lbdak{k}+\lbdak{k}\LAP_{\tsvk{k}}\left(s\right)}
} \nonumber \\
& \times\frac{\muk{k}^{\dit{i}{(c)}}r^{\dit{i}{(c)}-1}\exp\left(-\muk{k} r\right)}{\left(\dit{i}{(c)}-1\right)!}
dtdr
\end{align}
\blue{for $T_{\text{th},i}-r>0$}, where $\ilapt{\tquk{k}}$ is the inverse Laplace transform for \ac{pdf} of waiting time, \blue{$T_{\text{th},i}$ is $\ttgi{i}-\sum_{q\in\SetQ}\tcmwik{i}{k}$ for $q=\{\text{u, d}\}$}, and $\rhok{k}$ is the utilization factor of a $k$-tier \ac{MEC} server given by
\begin{align} \label{eq:rho}
\rhok{k} =\sum_{i\in\SetI}\lbdaik{i}{k}\frac{\dit{i}{(c)}}{\muk{k}}
\end{align}
for $0\leqq\rhok{k}<1$. In \eqref{eq:pcompikint}, $\LAP_{\tsvk{k}}\left(s\right)$ is the Laplace transform of the \ac{pdf} of service time in a $k$-tier \ac{MEC} server given by
\begin{align} \label{eq:tau}
\LAP_{\tsvk{k}}\left(s\right)=\sum_{i\in\SetI}\frac{\lbdaik{i}{k}}{\lbdak{k}}\left(\frac{\muk{k}}{s+\muk{k}}\right)^{\dit{i}{(c)}}.
\end{align}
\end{theorem}

\begin{IEEEproof}
The Laplace transform of the \ac{pdf} of waiting time is refered to as the Pollaczek-Khinchin (P-K) transform equation of M/G/1 queue in \cite{kle:75}. Using the equation, $\LAP_{\tquk{k}} (s)$ is given by 
\begin{align} \label{eq:applappk}
\LAP_{\tquk{k}}\left(s\right)
=\frac{\left(1-\rhok{k}\right)s}
{s-\lbdak{k}+\lbdak{k}\LAP_{\tsvk{k}}\left(s\right)}.
\end{align}
The $\ilapt{\tquk{k}}$ is obtained using \eqref{eq:applappk}, which shows the \ac{pdf} of the waiting time. Since $\tsvik{i}{k}$ is a random variable with the \ac{pdf} in \eqref{eq:svtikpdf}, $\pecik{i}{k}$ is given by
\begin{align} \label{eq:apppcompikint}
\pecik{i}{k} =& \int_{0}^{T_{\text{th},i}}\!\!\!\int_{0}^{T_{\text{th},i}-r}
\!\!\!\ilaptt{\tquk{k}}{
	\frac{\left(1-\rhok{k}\right)s}
	{ s-\lbdak{k}+\lbdak{k}\LAP_{\tsvk{k}}\left(s\right)}
} \nonumber \\
&\times\fsvikr{i}{k}
dtdr.
\end{align}
According to \cite{kle:75}, $\rhok{k}$ and $\LAP_{\tsvk{k}}\left(s\right)$ are given by, respectively
\begin{align} \label{eq:apprhok}
\rhok{k}&=\lbdak{k}\ex{\tsvk{k}}
=\lbdak{k}\sum_{i\in\SetI}\frac{\lbdaik{i}{k}}{\lbdak{k}}\frac{\dit{i}{(c)}}{\muk{k}}
=\sum_{i\in\SetI}\lbdaik{i}{k}\frac{\dit{i}{(c)}}{\muk{k}}
\end{align}
\vspace{-3mm}
\begin{align} \label{eq:apptsvkl}
\LAP_{\tsvk{k}}\left(s\right)
&=\sum_{i\in\SetI}\frac{\lbdaik{i}{k}}{\lbdak{k}}\LAP\left[\frac{\muk{k}^{\dit{i}{(c)}}t^{\dit{i}{(c)}-1}\exp\left(-\muk{k} t\right)}{\left(\dit{i}{(c)}-1\right)!}\right] \nonumber \\
&=\sum_{i\in\SetI}\frac{\lbdaik{i}{k}}{\lbdak{k}}\left(\frac{\muk{k}}{s+\muk{k}}\right)^{\dit{i}{(c)}}.
\end{align}
Substituting \eqref{eq:apprhok}, \eqref{eq:apptsvkl} and \eqref{eq:svtikpdf} into \eqref{eq:applappk}, \eqref{eq:apppcompikint} becomes \eqref{eq:pcompikint}. 
\end{IEEEproof}
The $\pecik{i}{k}$ is hard to be presented in a closed form because of the inverse Laplace transform. However, $\pecik{i}{k}$ can be given in a closed form for some cases as the following corollaries.
\setcounter{eqnback}{\value{equation}}
\setcounter{equation}{45}
\begin{figure*}[h!]
	\blue{
	\begin{align} 
	\pecikh{i}{k}
	=&
	\left(1-\rhok{k}\right)\!
	\frac{\gamma\left(\dit{i}{(c)},\muk{k}T_{\text{th},i}\right)}{\left(\dit{i}{(c)}-1\right)!}
	\!+\! 
	\rhok{k}
	\biggr[\frac{\gamma\left(\beta_{k,1},\beta_{k,2}T_{\text{th},i}\right)}{\Gamma\left(\beta_{k,1}\right)}\!-\!
	\frac{\beta_{k,2}^{\beta_{k,1}}\exp\{-\beta_{k,2}T_{\text{th},i}\}}{\Gamma\left(\beta_{k,1}\right)} \nonumber \\
	&\times\!\sum_{n=0}^{\dit{i}{(c)}-1}\!\!\frac{\muk{k}^{n}}{n!}
	B\left(\beta_{k,1},n+1\right)T_{\text{th},i}^{\beta_{k,1}+n}
	\hgf{1}{1}{n+1;\beta_{k,1}+n+1;T_{\text{th},i}\left(\beta_{k,2}-\muk{k}\right)}\biggr]
	\label{eq:cdf2123}
	\end{align}
	\setcounter{eqncnt}{46}
	\centering \rule[0pt]{18cm}{0.3pt}
	}
\end{figure*}
\setcounter{equation}{\value{eqnback}}

\begin{corollary} \label{coro:cdf21}
For the user type set $\SetI = \{1\}$ and $\dit{i}{(c)} = 1$, $\pecik{i}{k}$ is given by
\begin{align} \label{eq:cdf21}
\pecik{i}{k}
=
1-\exp\left\{\left(-\muk{k}\!+\!\lbdak{k}\right)T_{\text{th},i}\right\}.
\end{align}
\begin{IEEEproof}
For the user type set $\SetI = \{1\}$ and $\dit{i}{(c)} = 1$, $\LAP_{\tsvk{k}}\left(s\right) = \frac{\muk{k}}{s+\muk{k}}$, so we have 
%
\begin{align} \label{eq:applapqpdf21}
\LAP_{\tquk{k}}\!\!\left(s\right)
\!=\!\frac{\left(1-\rhok{k}\right)s}
{s-\lbdak{k}+\frac{\lbdak{k}\muk{k}}{s+\muk{k}}}
\!=\!\left(1\!-\rhok{k}\right)\!\left\{1\!+\!\frac{\lbdak{k}}{s\!+\!\muk{k}\!-\!\lbdak{k}}\!\right\}\!.
\end{align}
Since $\LAP^{-1}\{\frac{1}{s+\muk{k}-\lbdak{k}}\}=\exp\{\left(-\muk{k}+\lbdak{k}\right)t\}$, $f_{\tquk{k}}\!\left(t\right)\!$, i.e., $\ilapt{\tquk{k}}$, is presented by
\begin{align} \label{eq:apppdf21}
f_{\tquk{k}}\!\left(t\right)\!=\!\left(1-\rhok{k}\right)\!\delta\left(t\right)\!+\!\lbdak{k}\!\left(1-\rhok{k}\right)\exp\!\left\{\left(-\muk{k}\!+\!\lbdak{k}\right)\!t\right\}
\end{align}
where $\delta\left(t\right)$ is the delta function which means that a user has zero wait with probability $\left(1-\rhok{k}\right)$. Since $\tsvik{i}{k}$ in threshold is a random variable, by substituting  $\ilaptt{\tquk{k}}{\!\frac{\left(1-\rhok{k}\right)s}{s-\lbdak{k}+\lbdak{k}\LAP_{\tsvk{k}}\left(s\right)}\!}$ in \eqref{eq:pcompikint} into \eqref{eq:apppdf21} and applying the \ac{pdf} of $\tsvik{i}{k}$, $\pecik{i}{k}$ is given by 
\begin{align} \label{eq:apppcompik21}
\pecik{i}{k}
=&
\frac{\gamma\left(\dit{i}{(c)},\muk{k}T_{\text{th},i}\right)}
{\left(\dit{i}{(c)}-1\right)!}
-
\rhok{k}^{1-\dit{i}{(c)}}
\exp\left\{-\left(\muk{k}-\lbdak{k}\right)T_{\text{th},i}\right\} \nonumber \\
&\times\frac{
\gamma\left(\dit{i}{(c)},\lbdak{k}T_{\text{th},i}\right)
}
{\left(\dit{i}{(c)}-1\right)!}
\end{align}
Substituting $\dit{i}{(c)}$ into $1$, \eqref{eq:apppcompik21} becomes \eqref{eq:cdf21}.
\end{IEEEproof}
	
\end{corollary}

\begin{corollary} \label{coro:cdf212}
For the user type set $\SetI = \{1, 2\}$ and $\dit{i}{(c)} \in \{1, 2\}$, $\pecik{i}{k}$ is given by
\begin{align} \label{eq:cdf212}
\pecik{i}{k}
=&1-\rhok{k}+\frac{1-\rhok{k}}{\zeta_{1}-\zeta_{2}}\left[-\frac{\left(\zeta_{1}+\muk{k}\right)^{2}}{\zeta_{1}}\!+\frac{\left(\zeta_{2}+\muk{k}\right)^{2}}{\zeta_{2}}\!\right. \nonumber \\
&
\!+\frac{\left(\zeta_{1}+\muk{k}\right)^{2}}{\zeta_{1}}
\!
\left(\frac{\muk{k}}{\muk{k}+\zeta_{1}}\right)^{\!\!\dit{i}{(c)}}\!\!\!\!\!\!\!\exp\left\{\zeta_{1}T_{\text{th},i}\right\}\!\! \nonumber \\
&
\left.\!-\frac{\left(\zeta_{2}+\muk{k}\right)^{2}}{\zeta_{2}}
\!
\left(\frac{\muk{k}}{\muk{k}+\zeta_{2}}\right)^{\!\!\dit{i}{(c)}}\!\!\!\!\!\!\!\exp\left\{\zeta_{2}T_{\text{th},i}\right\}\!
\right]
\end{align}
where $\zeta_{1}$ and $\zeta_{2}$ are given, respectively, by
\begin{align} \label{eq:zeta1}
\zeta_{1}=-\muk{k}+\frac{\lbdak{k}}{2}+\sqrt{\frac{\lbdak{k}^{2}}{4}+\muk{k}\left(\lbdak{k}-\lbdaik{1}{k}\right)}, \nonumber \\
\zeta_{2}=-\muk{k}+\frac{\lbdak{k}}{2}-\sqrt{\frac{\lbdak{k}^{2}}{4}+\muk{k}\left(\lbdak{k}-\lbdaik{1}{k}\right)}.
\end{align}
\vspace{-4mm}
\begin{IEEEproof}
For the user type set $\SetI = \{1, 2\}$ and $\dit{i}{(c)} \in \{1, 2\}$, $\LAP_{\tsvk{k}}\left(s\right)=\frac{\lbdaik{1}{k}}{\lbdak{k}}\frac{\muk{k}}{s+\muk{k}}+\frac{\lbdaik{2}{k}}{\lbdak{k}}\left(\frac{\muk{k}}{s+\muk{k}}\right)^{2}$, so $\LAP_{\tquk{k}}\left(s\right)$ is given by
\begin{align} \label{eq:applapqpdf212}
\LAP_{\tquk{k}}\left(s\right)
=&\frac{\left(1-\rhok{k}\right)s}
{s-\lbdak{k}+\lbdak{k}
\left(\frac{\lbdaik{1}{k}}{\lbdak{k}}\frac{\muk{k}}{s+\muk{k}}+\frac{\lbdaik{2}{k}}{\lbdak{k}}\left(\frac{\muk{k}}{s+\muk{k}}\right)^{2}\right)}\nonumber \\
=&\frac{1-\rhok{k}}{\zeta_{1}-\zeta_{2}}\!
\!\left\{\!\frac{\left(\zeta_{1}\!+\!\muk{k}\right)^{2}}{s-\zeta_{1}}
\!-\!\frac{\left(\zeta_{2}\!+\!\muk{k}\right)^{2}}{s-\zeta_{2}}
\!+\zeta_{1}\!-\zeta_{2}\!\right\}
\end{align}
where $\zeta_{1}$ and $\zeta_{2}$ are \eqref{eq:zeta1}, and $\lbdak{k}=\lbdaik{1}{k}+\lbdaik{2}{k}$. Since $\LAP^{-1}\{\frac{1}{s-\zeta_{1}}\}=\exp\{\zeta_{1}t\}$, $f_{\tquk{k}}\!\left(t\right)\!$, i.e., $\ilapt{\tquk{k}}$, is given by 
\begin{align} \label{eq:apppdf212}
f_{\tquk{k}}\left(t\right)
=&\left(1-\rhok{k}\right)\delta\left(t\right)+\frac{1-\rhok{k}}{\zeta_{1}-\zeta_{2}} 
\left[\left(\zeta_{1}+\muk{k}\right)^{2}\exp\left\{\zeta_{1}t\right\}\right. \nonumber \\
&\left.-\left(\zeta_{2}+\muk{k}\right)^{2}\exp\left\{\zeta_{2}t\right\}\right].
\end{align}
Substituting $\ilaptt{\tquk{k}}{\!\frac{\left(1-\rhok{k}\right)s}{s-\lbdak{k}+\lbdak{k}\LAP_{\tsvk{k}}\left(s\right)}\!}$ in \eqref{eq:pcompikint} into \eqref{eq:apppdf212}, $\pecik{i}{k}$ is given by
\begin{align} \label{eq:apppcompik212}
\pecik{i}{k}
=&\int_{0}^{\infty}
\!\biggr[\frac{1-\rhok{k}}{\zeta_{1}-\zeta_{2}} 
\biggr\{\frac{\left(\zeta_{1}\!+\!\muk{k}\right)^{2}}{\zeta_{1}}
\left(\exp\{\zeta_{1}\left(T_{\text{th},i}-r\right)\}\!-\!1\right) \nonumber \\
&\!\!-\!\frac{\left(\zeta_{2}\!+\!\muk{k}\right)^{2}}{\zeta_{2}}\!\left(\exp\{\zeta_{2}\left(T_{\text{th},i}-r\right)\}\!-\!1\right)\!\biggr\}\!+\!\left(1-\rhok{k}\right)\biggr] \nonumber \\
&\times\frac{\muk{k}^{\dit{i}{(c)}}r^{\dit{i}{(c)}-1}\!\exp\!\left(-\muk{k} r\right)}{\left(\dit{i}{(c)}-1\right)!}
dr
\end{align}
which is the same as \eqref{eq:cdf212}.
\end{IEEEproof}
\end{corollary}

\begin{figure}[t!]
	\begin{center}   
		{ 
			\psfrag{pcomp}[bc][tc][0.7]{$\pec$, $\pech$}
			\psfrag{targetlatency}[tc][bc][0.7]{$\ttgi{i}$ [sec]}
			\psfrag{pcomp-21}[bl][bl][0.59]{$\pecik{i}{k}$ for Corollary \ref{coro:cdf21}}	
			\psfrag{pcomp-212-pui-0.5,0.5}[bl][bl][0.59]{$\pecik{i}{k}$ for Corollary \ref{coro:cdf212}}
			\psfrag{pcomp-2123-ga-pui-0.4,0.2,0.4}[bl][bl][0.59]{$\pecikh{i}{k}$ for Lemma \ref{coro:cdf2123}}
			\psfrag{pcomp-2123-pui-0.4,0.2,0.4-5-00000}[bl][bl][0.59]{$\pecik{i}{k}$ for Lemma \ref{coro:cdf2123} (Simulation)}
			\includegraphics[width=1.00\columnwidth]{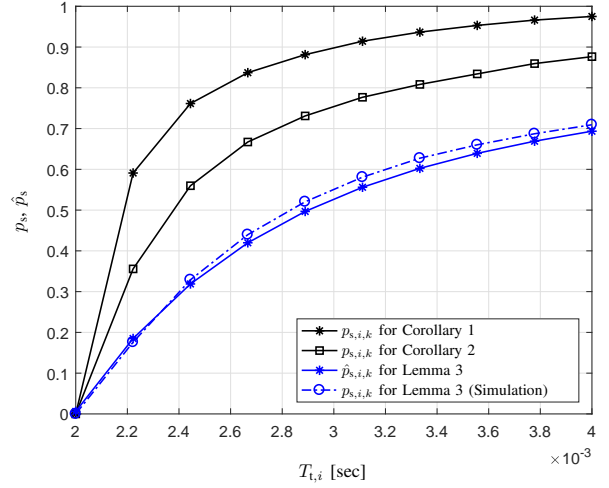}
		}
	\end{center}
	\vspace{-2mm}
	\caption{
		\ac{SECP} $\pec$ and approximated \ac{SECP} $\pech$ in $2$ tier \ac{MEC}-enabled \ac{HetNet} as a function of the target latency $\ttgi{i}$ for different number of user type. 
	}
	\vspace{0mm}
	\label{fig:cp}
\end{figure}

\blue{
For general cases, we can describe $\pecikh{i}{k}$ as a closed form by approximating the waiting time distribution via Gamma distribution \cite{menhenzeptra:06} to decrease the computational complexity. The approximated \ac{SECP}, denoted by $\pecikh{i}{k}$, is presented in the following lemma
\begin{lemma} \label{coro:cdf2123}
For every user type set $\SetI$ and $\dit{i}{(c)}$, the approximated \ac{SECP} of an $i$-type user offloading to a $k$-tier \ac{MEC} server, $\pecikh{i}{k}$, is given by \eqref{eq:cdf2123},
where $\hgf{1}{1}{\cdot;\cdot;\cdot}$ is the confluent hypergeometric function of the first kind, and $B\left(\cdot,\cdot\right)$ is the beta function.
In \eqref{eq:cdf2123}, $\beta_{1}$ and $\beta_{2}$ are defined, respectively, as
\setcounter{equation}{46}
\begin{align} \label{eq:b1}
\beta_{k,1} = \frac{\ex{\tquk{k}}^{2}}{\ex{\tquk{k}^{2}}-\ex{\tquk{k}}^{2}},\indent\beta_{k,2} = \frac{\beta_{k,1}}{\ex{\tquk{k}}}
\end{align}
where $\ex{\tquk{k}}$ and $\ex{\tquk{k}^{2}}$ are given, respectively, by
\begin{align} \label{eq:tqukga}
\ex{\tquk{k}}=
\frac{\sum_{i\in\SetI}\lbdaik{i}{k}\dit{i}{(c)}\left(\dit{i}{(c)}+1\right)}
{2\muk{k}^{2}\left(1-\sum_{i\in\SetI}\lbdaik{i}{k}\frac{\dit{i}{(c)}}{\muk{k}}\right)}
\end{align}
\begin{align} \label{eq:tquksqga}
\ex{\tquk{k}^{2}}\!=\!
2\ex{\tquk{k}}^{2}\!+\!
\frac{\sum_{i\in\SetI}\lbdaik{i}{k}\dit{i}{(c)}\left(\dit{i}{(c)}+1\right)\!\left(\dit{i}{(c)}+2\right)}
{3\muk{k}^{3}\left(1\!-\!\sum_{i\in\SetI}\lbdaik{i}{k}\frac{\dit{i}{(c)}}{\muk{k}}\right)}.
\end{align}
\begin{IEEEproof}
	See Appendix~\ref{app:corocdf2123}. 
\end{IEEEproof}
\end{lemma}
}
\begin{table}[!t] 
	\caption{Parameter values if not otherwise specified} 
	\vspace{-4mm}
	\begin{center}
		\rowcolors{2}
		{cyan!15!}{}
		\renewcommand{\arraystretch}{1.3}
		\begin{tabular}{l l | l l}
			\hline 
			{\bf Parameters} & {\bf Values} & {\bf Parameters} & {\hspace{0.32cm}}{\bf Values} \\
			\hline
			\hspace{0.15cm}$\lbdm$ [nodes/m$^2$] & \hspace{0.2cm}$4*10^{-5}$ 
			& \hspace{0.12cm}$\lbdu$ [nodes/m$^2$] & \hspace{0.2cm}$12*10^{-4}$ \\ 
			\hspace{0.15cm}$\pui{1}$ & \hspace{0.2cm}$0.5$ 
			& \hspace{0.12cm}$\pui{2}$ & \hspace{0.2cm}$0.5$  \\ 
			\hspace{0.15cm}$\pmk{1}$ & \hspace{0.2cm}$0.25$ 
			& \hspace{0.12cm}$\pmk{2}$ & \hspace{0.2cm}$0.75$  \\ 
			\hspace{0.15cm}$\bw{(u)}$ [Hz] & \hspace{0.2cm}$5*10^{6}$ 
			& \hspace{0.12cm}$\bw{(d)}$ [Hz] & \hspace{0.2cm}$10*10^{6}$  \\ 
			\hspace{0.15cm}$\pwrui{i}$ [dBm] & \hspace{0.2cm}$23$ 
			& \hspace{0.12cm}$\pwrmk{1}$ [dBm] & \hspace{0.2cm}$43$  \\ 
			\hspace{0.15cm}$\pwrmk{2}$ [dBm] & \hspace{0.2cm}$33$ 
			& \hspace{0.12cm}$N_{0}$ [dBm] & \hspace{0.2cm}$-104$ \\ 
			\hspace{0.15cm}$\alpha$ & \hspace{0.2cm}$4$ 
			& \hspace{0.12cm}$\kappa$ & \hspace{0.2cm}$0.75$ \\ 
			\hspace{0.15cm}$\betauik{i}{k}$ & \hspace{0.2cm}$0.95$ 
			& \hspace{0.12cm}$\betadik{i}{k}$ & \hspace{0.2cm}$0.95$  \\ 
			\hspace{0.15cm}$\muk{1}$ [packet/slot] & \hspace{0.2cm}$9$  
			& \hspace{0.12cm}$\muk{2}$ [packet/slot]& \hspace{0.2cm}$3$  \\ 
			\hspace{0.15cm}$\punit$ [KB] & \hspace{0.2cm}$100$
			& \hspace{0.12cm}$C_{\text{u}}$ [cycles/bit] & \hspace{0.2cm}$1400$  \\ 
			\hspace{0.15cm}$F_{\text{m},1}$ [Hz] & \hspace{0.2cm}$10*10^{6}$  
			& \hspace{0.12cm}$F_{\text{m},2}$ [Hz]& \hspace{0.2cm}$3*10^{6}$  \\ 
			\hspace{0.15cm}$\ts$ [sec] &  \hspace{0.2cm}$10^{-3}$
			& \hspace{0.12cm} & \hspace{0.2cm}  \\ 
			\hline
		\end{tabular}		 \vspace{-2mm}
	\end{center}
	\label{table:simpar}
\end{table}

Fig. \ref{fig:cp} shows the \acp{SECP} of the corollaries, and the approximated \ac{SECP} of the lemma as a function of $\ttgi{i}$ for fixed $\poik{i}{k}$ considering the $2$ tier \ac{MEC}-enabled \ac{HetNet}.
\blue{For this figure, $\ttgi{i}$ are set to be equal for all types of users, and other parameters in Table \ref{table:simpar} are used. From Fig. \ref{fig:cp}, we can see the good match between $\pech$ for Lemma \ref{coro:cdf2123} and $\pec$ obtained by the simulation for Lemma \ref{coro:cdf2123}.} Hence, the results of Lemma \ref{coro:cdf2123} can be used to get the numerical results.

Substituting \eqref{eq:poik}, and \eqref{eq:pcompikint} into \eqref{eq:ps}, $\pec$ can be derived. From \eqref{eq:ps}, we can see that deriving the optimal bias factors in terms of $\pec$ is hard due to the complex structure. However, we show the existence of the optimal bias factor using numerical results in Section \ref{sec:numresult}.

\section{Numerical Results} \label{sec:numresult}

\blue{
In this section, we provide numerical results on the \ac{SECP} for the \ac{MEC}-enabled \ac{HetNet} consisted of $2$ tier networks (except for Fig. \ref{fig:secp3tier} that considers $3$ tier networks).
The $1$-tier \ac{MEC} servers have higher computing capabilities than the $2$-tier \ac{MEC} servers, and computing capabilities for the $2$-tier \ac{MEC} servers are also higher than the $3$-tier \ac{MEC} servers.
We assume that both $\betauik{i}{k}$ and $\betadik{i}{k}$ are set to be equal.
Note that our framework can be easily extended to the network with different values of $\betauik{i}{k}$ and $\betadik{i}{k}$ by simply changing the parameters.
The values of parameters used for numerical results are given in Table \ref{table:simpar} according to \cite{lopguvdekouquezha:11,dammonweijiluovajyoosonmal:11,youhuachakim:17,chejialifu:15,liubenpoo:17}.
The other parameters not presented in Table \ref{table:simpar} are mentioned when the corresponding figures are introduced.
}

\subsection{\blue{SECP - Impact of Network Parameters}}

\begin{figure}[t!]
	\begin{center}   
		{ 
			\psfrag{pspcomp}[bc][tc][0.8]{$\pec$, $\pcomp$}
			\psfrag{bdb}[tc][bc][0.7]{$\biasik{1}{2}$ [dB]}
			\psfrag{secp-analysis}[bl][bl][0.59]{Analysis $\pec$}	
			\psfrag{scp-analysis}[bl][bl][0.59]{Analysis $\pcomp$}
			\psfrag{secp-simulation}[bl][bl][0.59]{Simulation $\pec$}
			\psfrag{scp-simulation}[bl][bl][0.59]{Simulation $\pcomp$}
			\includegraphics[width=1.00\columnwidth]{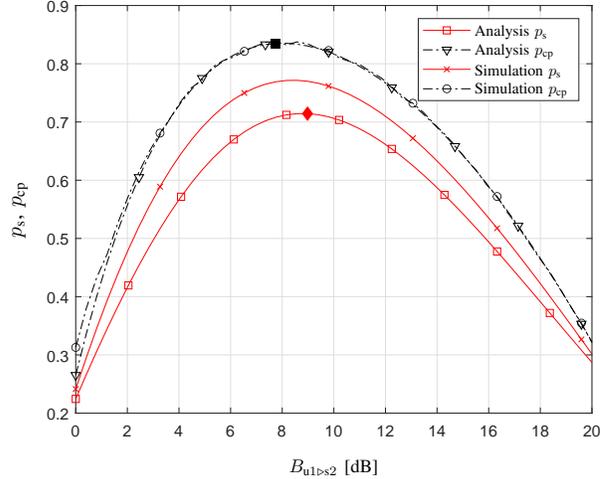}
		}
	\end{center}
	\vspace{-2mm}
	\caption{
		\ac{SECP} $\pec$ and \ac{SCP} $\pcomp$ 
		of $2$ tier \ac{MEC}-enabled \ac{HetNet} with $1$ user type
		as a function of bias factor of a $1$-type user to a $2$-tier \ac{MEC} server $\biasik{1}{2}$.		
	}
	\vspace{0mm}
	\label{fig:secpscp21sim}
\end{figure}

\blue{
In this subsection, we show how the bias factors in \ac{MEC} server association affect the \ac{SECP} for different network parameters via numerical results. To evaluate the \ac{SECP}, we first define the \ac{SCP} as the probability that the computation in \ac{MEC} server is finished within a target latency. The \ac{SCP} for an $i$-type user offloading to a $k$-tier \ac{MEC} server, denoted by $\pcpik{i}{k}$, is defined by $\pcpik{i}{k}=\px{\tquk{k}+\tsvik{i}{k}\leqq\ttgi{i}}$. The overall \ac{SCP} $\pcp$ is presented by $\pcp=\sum_{k\in\SetK}\sum_{i\in\SetI}\pui{i}\poik{i}{k}\pcpik{i}{k}$.
}

\blue{
Fig. \ref{fig:secpscp21sim} shows $\pec$ and $\pcomp$ with $1$ type of users (i.e., $\SetI = \{1\}$ with $\dit{1}{(c)}=\{1\}$) as a function of the bias factor $\biasik{1}{2}$ when $\ttgi{i}=1\times10^{-3}$s and $\biasik{1}{1} = 10\text{dB}$.
Note that increasing $\biasik{1}{2}$ (i.e., x-axis in Fig. \ref{fig:secpscp21sim}) means more users offload their required computation to a $2$-tier \ac{MEC} server than to a $1$-tier \ac{MEC} server.
From Fig. \ref{fig:secpscp21sim}, we first see that the simulation results of $\pcomp$ matches well with our analysis, while that of $\pec$ does not match well due to the assumptions used in the communication latency analysis.
However, we can still see that the trends according to the bias factor are the same.
}

\blue{
In addition, from Fig. \ref{fig:secpscp21sim}, it can be seen that when $\biasik{1}{2}$ is small, both $\pcomp$ and $\pec$ are small.
This is because the $1$-tier \ac{MEC} servers are heavy-loaded, and the communication link between a user and the \ac{MEC} server are long due to the lower \ac{MEC} server density in $1$-tier $\lbdmk{1}$ than that of $2$-tier server $\lbdmk{2}$.
On the other hands, as $\biasik{1}{2}$ increases, both $\pcomp$ and $\pec$ increase because the computation tasks are starting to be offloaded to a $2$-tier \ac{MEC} server, which can be located closer to the users and less loaded.
However, after certain points of $\biasik{1}{2}$, both $\pcomp$ and $\pec$ decrease since a $2$-tier \ac{MEC} server becomes heavy-loaded, and the communication link becomes longer.
Therefore, there exists the optimal bias factor $\biasoptik{1}{2}$ (marked by filled diamond marker for $\pec$ and filled rectangle marker for $\pcomp$) in terms of $\pcomp$ and $\pec$, which are different.
}

\blue{
In general \acp{HetNet}, the optimal bias factor is determined to offload more task to the $2$-tier servers which have the advantage of shorter link distance.
On the contrary, when we consider the computing capability of \ac{MEC} servers and the amount of computation tasks, the optimal bias factor for $\pcomp$ is located closer to zero to offload to the high-capable \ac{MEC} servers.
Therefore, $\biasoptik{i}{k}$ for $\pec$ are larger than $\biasoptik{i}{k}$ for $\pcomp$ as shown in Fig. \ref{fig:secpscp21sim}.
In other words, the \ac{MEC}-enabled \acp{HetNet} has different optimal bias factors from conventional ones, which consider the communication performance only or computation performance only.
Specifically, the optimal bias factor is located between the optimal bias factors obtained in terms of computing or communication performance only.
}

\begin{figure}
	\begin{center}   
		{ 
			\psfrag{secp}[bc][tc][0.8]{$\pec$}
			\psfrag{scp}[bc][tc][0.8]{$\pcomp$}
			\psfrag{bdb}[tc][bc][0.7]{$\biasik{2}{2}$ [dB]}	
			\psfrag{secp-reference}[bl][bl][0.59]{$\muk{2}=3$}
			\psfrag{secp-mu-low}[bl][bl][0.59]{$\muk{2}=3.45$}
			\psfrag{secp-mu-high}[bl][bl][0.59]{$\muk{2}=3.9$}
			\psfrag{scp-reference}[bl][bl][0.59]{$\muk{2}=3$}
			\psfrag{scp-mu-low}[bl][bl][0.59]{$\muk{2}=3.45$}
			\psfrag{scp-mu-high}[bl][bl][0.59]{$\muk{2}=3.9$}
			\includegraphics[width=1.00\columnwidth]{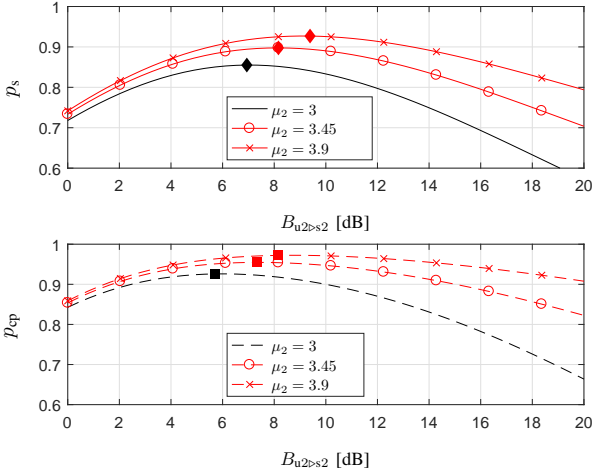}
		}
	\end{center}
	\vspace{-2mm}
	\caption{
		\ac{SECP} $\pec$
		and \ac{SCP} $\pcomp$
		of $2$ tier \ac{MEC}-enabled \ac{HetNet} with $2$ user type
		as a function of bias factor of a $2$-type user to a $2$-tier \ac{MEC} server $\biasik{2}{2}$
		for different service rates (i.e., computing capabilities) of a $2$-tier server $\muk{2}$.	
	}
	\vspace{0mm}
	\label{fig:secpscp22mu}
\end{figure}
\begin{figure}
	\begin{center}   
		{ 
			\psfrag{secp}[bc][tc][0.8]{$\pec$}
			\psfrag{scp}[bc][tc][0.8]{$\pcomp$}
			\psfrag{bdb}[tc][bc][0.7]{$\biasik{2}{2}$ [dB]}	
			\psfrag{secp-lbdu-low}[bl][bl][0.59]{$\lbdu=9.6*10^{-5}$}
			\psfrag{secp-lbdu-high}[bl][bl][0.59]{$\lbdu=10.8*10^{-5}$}
			\psfrag{secp-reference--------}[bl][bl][0.59]{$\lbdu=12*10^{-5}$}
			\psfrag{scp-lbdu-low}[bl][bl][0.59]{$\lbdu=9.6*10^{-5}$}
			\psfrag{scp-lbdu-high}[bl][bl][0.59]{$\lbdu=10.8*10^{-5}$}
			\psfrag{scp-reference--------}[bl][bl][0.59]{$\lbdu=12*10^{-5}$}
			\includegraphics[width=1.00\columnwidth]{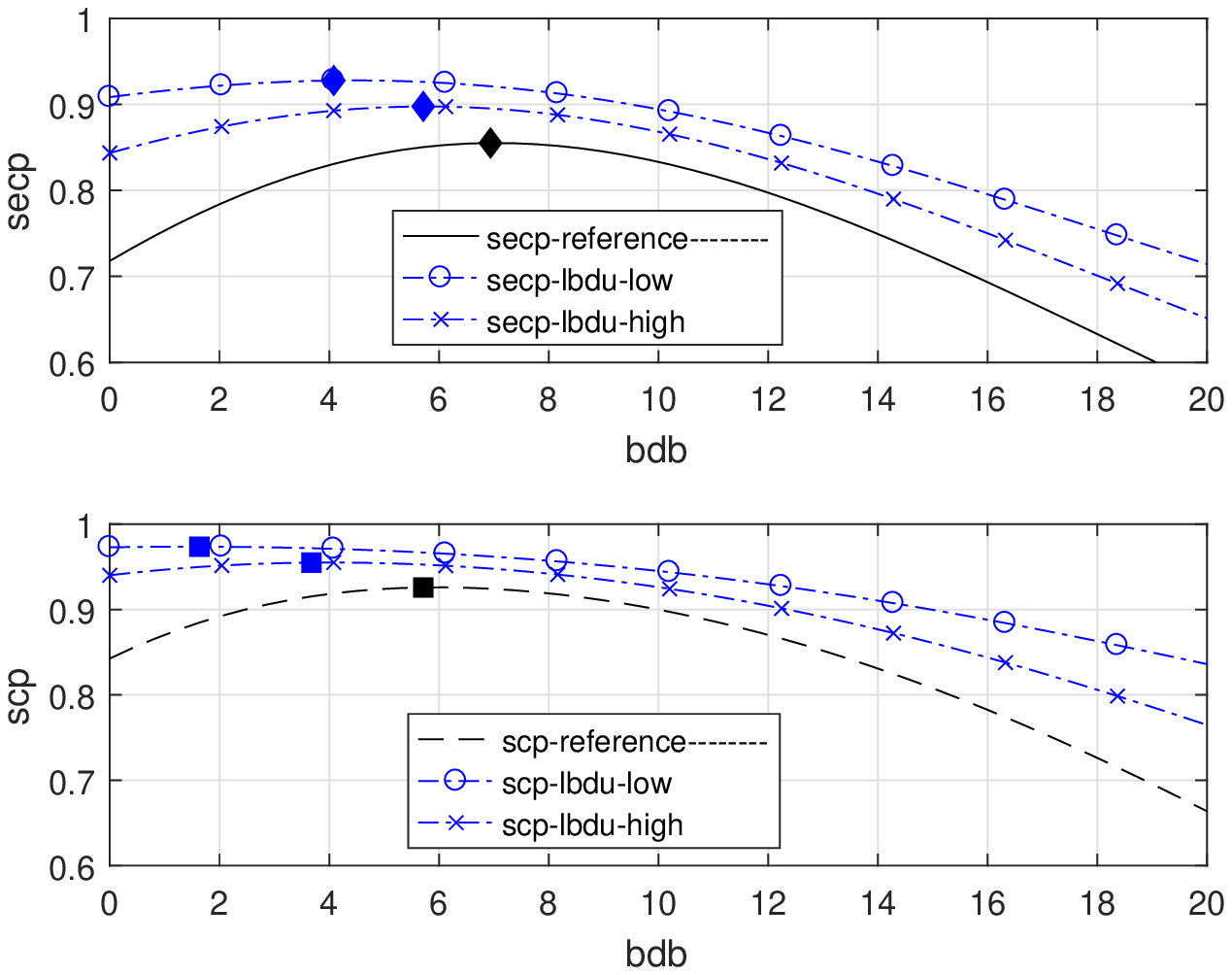}
		}
	\end{center}
	\vspace{-2mm}
	\caption{
		\ac{SECP} $\pec$
		and \ac{SCP} $\pcomp$
		of $2$ tier \ac{MEC}-enabled \ac{HetNet} with $2$ user type
		as a function of bias factor of a $2$-type user to a $2$-tier \ac{MEC} server $\biasik{2}{2}$
		for different user densities $\lbdu$.		
	}
	\vspace{0mm}
	\label{fig:secpscp22lbdu}
\end{figure}
\begin{figure}
	\begin{center}   
		{ 
			\psfrag{secpscp}[bc][tc][0.8]{$\pec$, $\pcomp$}
			\psfrag{bdb}[tc][bc][0.7]{$\biasik{2}{2}$ [dB]}	
			\psfrag{secp-dit-low}[bl][bl][0.59]{$\pec \text{ } (\dit{2}{(c)}=2)$}
			\psfrag{secp-dit-high}[bl][bl][0.59]{$\pec \text{ } (\dit{2}{(c)}=6)$}
			\psfrag{secp-reference}[bl][bl][0.59]{$\pec \text{ } (\dit{2}{(c)}=4)$}
			\psfrag{scp-dit-low}[bl][bl][0.59]{$\pcomp \text{ } (\dit{2}{(c)}=2)$}
			\psfrag{scp-dit-high}[bl][bl][0.59]{$\pcomp \text{ } (\dit{2}{(c)}=6)$}
			\psfrag{scp-reference}[bl][bl][0.59]{$\pcomp \text{ } (\dit{2}{(c)}=4)$}
			\psfrag{pcomp}[bl][bl][0.59]{$\pec=0.59$}
			\psfrag{designpcomp}[bl][bl][0.59]{$\biasoptik{2}{2} \text{ for } \pcomp$}
			\psfrag{ps}[bl][bl][0.59]{$\pec=0.63$}
			\psfrag{designps}[bl][bl][0.59]{$\biasoptik{2}{2} \text{ for } \pec$}
			\includegraphics[width=1.00\columnwidth]{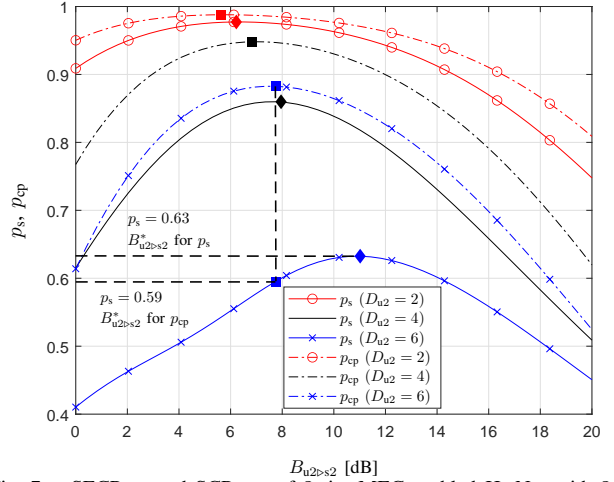}
		}
	\end{center}
	\vspace{-5mm}
	\caption{
		\ac{SECP} $\pec$
		and \ac{SCP} $\pcomp$
		of $2$ tier \ac{MEC}-enabled \ac{HetNet} with $2$ user type
		as a function of bias factor of a $2$-type user to a $2$-tier \ac{MEC} server $\biasik{2}{2}$
		for different task sizes of a $2$-type user.		
	}
	\vspace{-5mm}
	\label{fig:secpscp22dit}
\end{figure}
%
%

\blue{
Fig. \ref{fig:secpscp22mu} shows $\pec$ and $\pcomp$, respectively, with $2$ types of users (i.e., $\SetI = \{1,2\}$ with $\dit{i}{(c)}=\{1,2\}$) as a function of $\biasik{2}{2}$ for different computing capabilities of a $2$-tier \ac{MEC} server $\muk{2}$. 
For this figure, we consider $\ttgi{i}=2\times10^{-3}$s, and $\biasik{1}{2} = 10$dB.
We first see that both optimal bias factors $\biasoptik{2}{2}$ for $\pec$ and $\pcomp$ increase as $\muk{2}$ increases, since the $2$-tier \ac{MEC} servers can process the more computation tasks.
However, even for large $\muk{2}$, both $\pec$ and $\pcomp$ decrease as $\biasik{2}{2}$ increase after $\biasoptik{2}{2}$ due to the heavy-loaded \ac{MEC} servers.
Therefore, offloading more computation tasks to the high-capable \ac{MEC} servers generally shows better \ac{SECP} unless those servers are heavy-loaded.
Moreover, from Fig. \ref{fig:secpscp22mu}, it can be seen that $\biasoptik{2}{2}$ for $\pcomp$ are smaller than $\biasoptik{2}{2}$ for $\pec$.
As shown in Fig. \ref{fig:secpscp21sim}, $\pcomp$ only considers the computing capabilities, while $\pec$ considers the both short link distance and high computing capabilities.
}

\blue{
Figs. \ref{fig:secpscp22lbdu} and \ref{fig:secpscp22dit} present $\pec$ and $\pcomp$ as a function of $\biasik{2}{2}$ for different user densities $\lbdu$ and computation task sizes of a $2$-type user $\dit{2}{(c)}$, respectively, under the same environment of Fig. \ref{fig:secpscp22mu}.
We can see that as $\lbdu$ and $\dit{2}{(c)}$ increase, both $\biasoptik{2}{2}$ for $\pec$ and $\pcomp$ increase.
For small $\lbdu$ and $\dit{2}{(c)}$, $\biasoptik{2}{2}$ for both $\pec$ and $\pcomp$ become smaller because offloading to high-capable servers can achieves higher performance.
However, as $\lbdu$ and $\dit{2}{(c)}$ increase, offloading to the $1$-tier \ac{MEC} servers no longer enhances $\pec$ and $\pcomp$ due to the heavy-loaded $1$-tier servers.
Hence, $\biasoptik{2}{2}$ become larger to distribute the tasks to $2$-tier \ac{MEC} servers.
In Fig. \ref{fig:secpscp22dit}, we can also see that $\biasoptik{2}{2}$ for $\pec$ increase even more to not only distribute the tasks, but decrease the link distance increased by $\dit{2}{(c)}$.
Therefore, when the amount of computation tasks of the network is large, it is better to distribute the arrival of tasks to the low-capable \ac{MEC} servers instead of offloading the most of tasks to the high-capable servers.
}

\blue{
In Fig. \ref{fig:secpscp22dit}, it can be seen that as $\dit{i}{(c)}$ increases, the difference between $\pec$ when $\biasik{2}{2}$ is optimized for $\pec$ and $\pec$ when $\biasik{2}{2}$ is optimized for $\pcomp$ becomes larger.
Specifically, when $\dit{2}{(c)}=2$, $\pec$ when $\biasoptik{2}{2}=6.2$dB (i.e., optimized for $\pec$) is obtained as $0.98$, and $\pec$ when $\biasoptik{2}{2}=5.6$dB (i.e., optimized for $\pcomp$) is obtained as $0.97$.
While, when $\dit{2}{(c)}=6$, $\pec$ when $\biasoptik{2}{2}=11$dB (i.e., optimized for $\pec$) is given as $0.63$, and $\pec$ when $\biasoptik{2}{2}=7.7$dB (i.e., optimized for $\pcomp$) is given as $0.59$.
This implies that when designing the \ac{MEC}-enabled \ac{HetNet}, a bias design using \ac{SECP} can be more efficient than the conventional design using \ac{SCP}.
}



%
\begin{figure}
	\begin{center}   
		{ 
			\psfrag{bdb11}[tc][bc][0.7]{$\biasik{1}{1}$ [dB]}
			\psfrag{bdb12}[bc][tc][0.7]{$\biasik{1}{2}$ [dB]}
			\psfrag{pscp}[tc][tc][0.7]{$\pec$}
			\includegraphics[width=1.00\columnwidth]{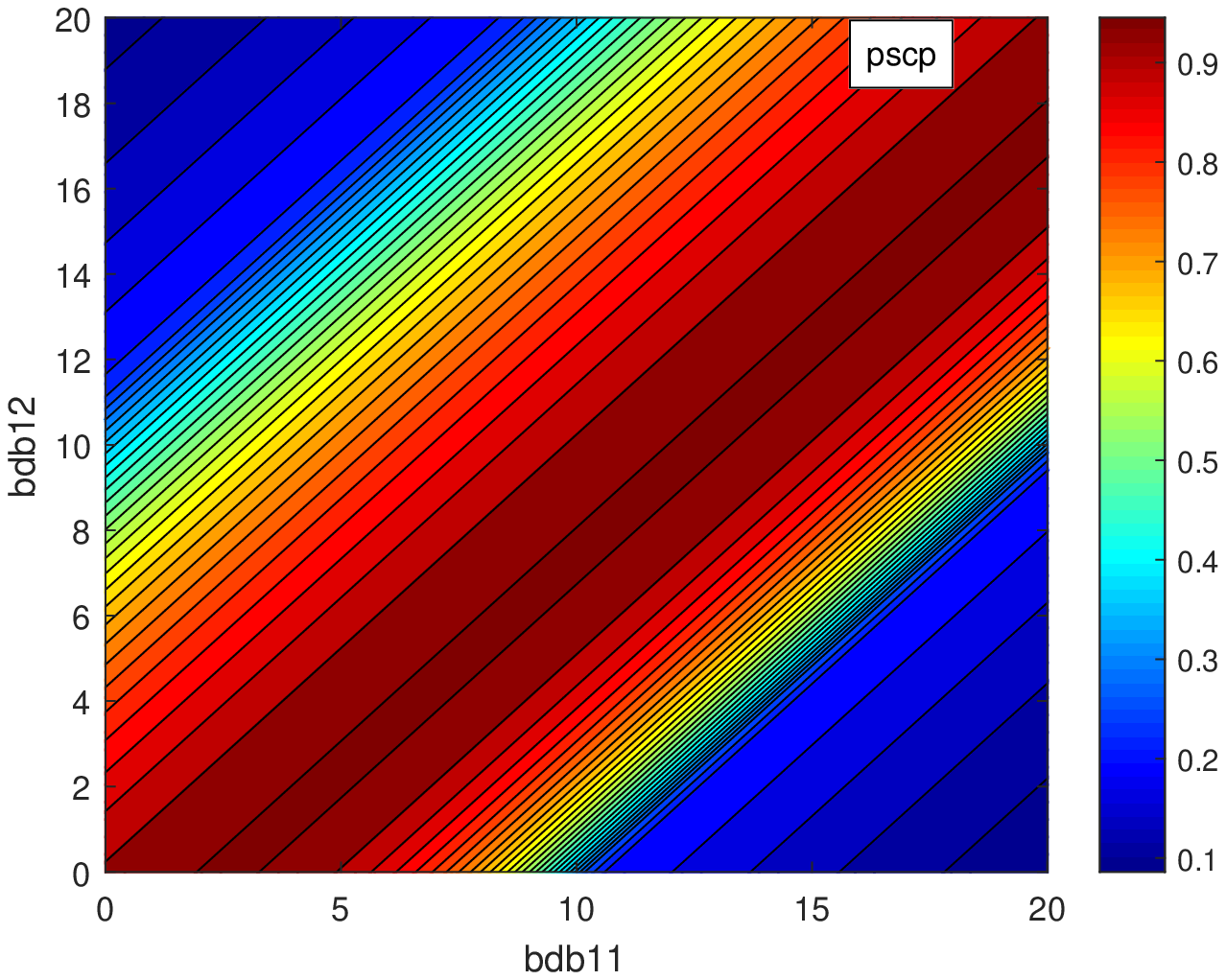}
		}
	\end{center}
	\vspace{-2mm}
	\caption{
		\ac{SECP} $\pec$
		of $2$ tier \ac{MEC}-enabled \ac{HetNet} with $1$ user type
		as a function of bias factor of a $1$-type user to $1$-tier \ac{MEC} server $\biasik{1}{1}$
		and bias factor of a $1$-type user to $2$-tier \ac{MEC} server $\biasik{1}{2}$.	
	}
	\vspace{0mm}
	\label{fig:secp2tier}
\end{figure}

\begin{figure}
	\begin{center}   
		{ 
			\psfrag{bdb12}[tc][bc][0.7]{$\biasik{1}{2}$ [dB]}
			\psfrag{bdb13}[bc][tc][0.7]{$\biasik{1}{3}$ [dB]}
			\psfrag{pscp}[tc][tc][0.7]{$\pec$}
			\includegraphics[width=1.00\columnwidth]{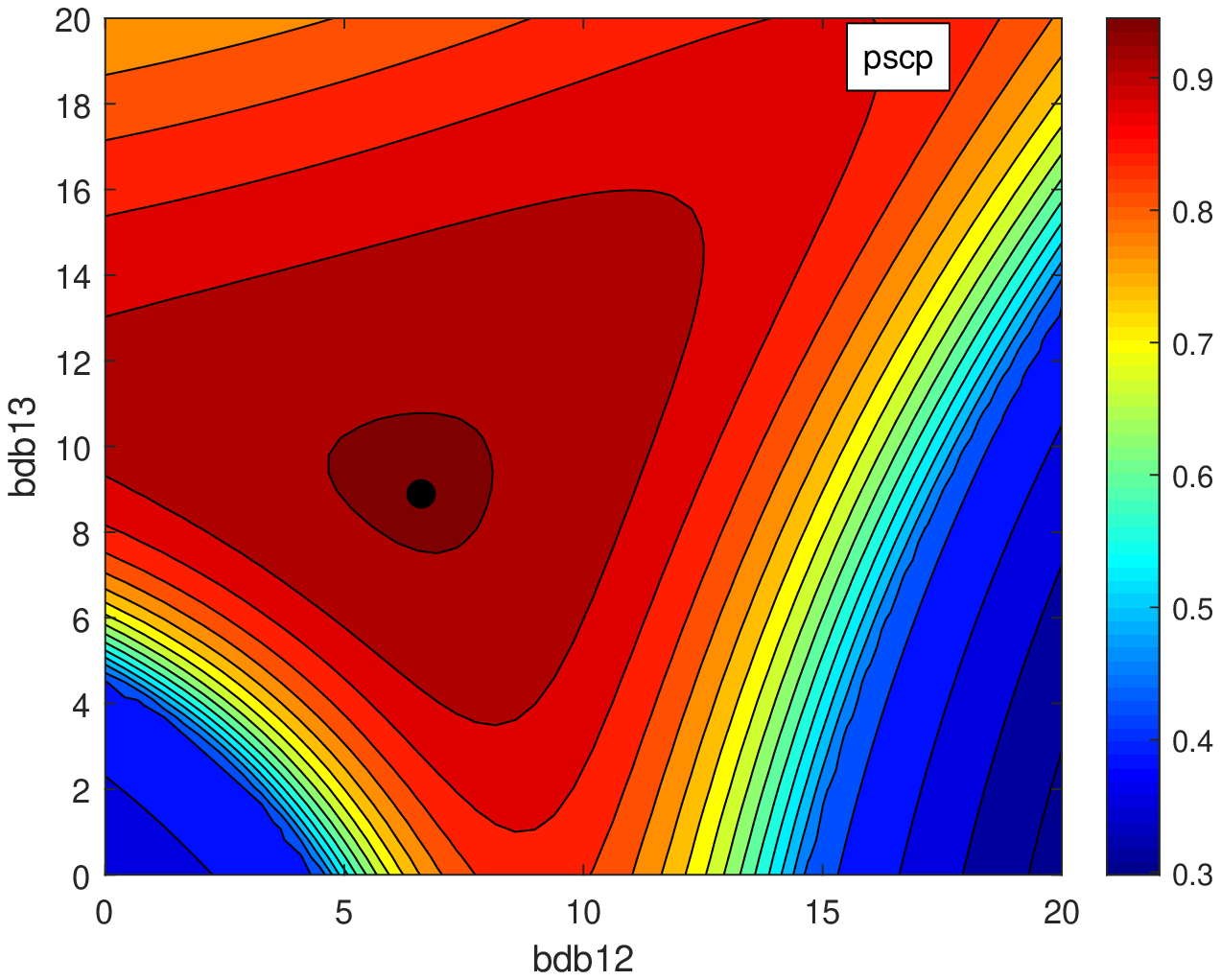}
		}
	\end{center}
	\vspace{-2mm}
	\caption{
		\ac{SECP} $\pec$
		of $3$ tier \ac{MEC}-enabled \ac{HetNet} with $1$ type user
		as a function of bias factor of a $1$-type user to $3$-tier \ac{MEC} server $\biasik{1}{2}$
		and bias factor of a $1$-type user to $3$-tier \ac{MEC} server $\biasik{1}{3}$.	
	}
	\vspace{0mm}
	\label{fig:secp3tier}
\end{figure}

\begin{figure}
	\begin{center}   
		{ 
			\psfrag{bdb12}[tc][bc][0.7]{$\biasik{1}{2}$ [dB]}
			\psfrag{bdb22}[bc][tc][0.7]{$\biasik{2}{2}$ [dB]}
			\psfrag{pscp}[tc][tc][0.7]{$\pcomp$}
			\includegraphics[width=1.00\columnwidth]{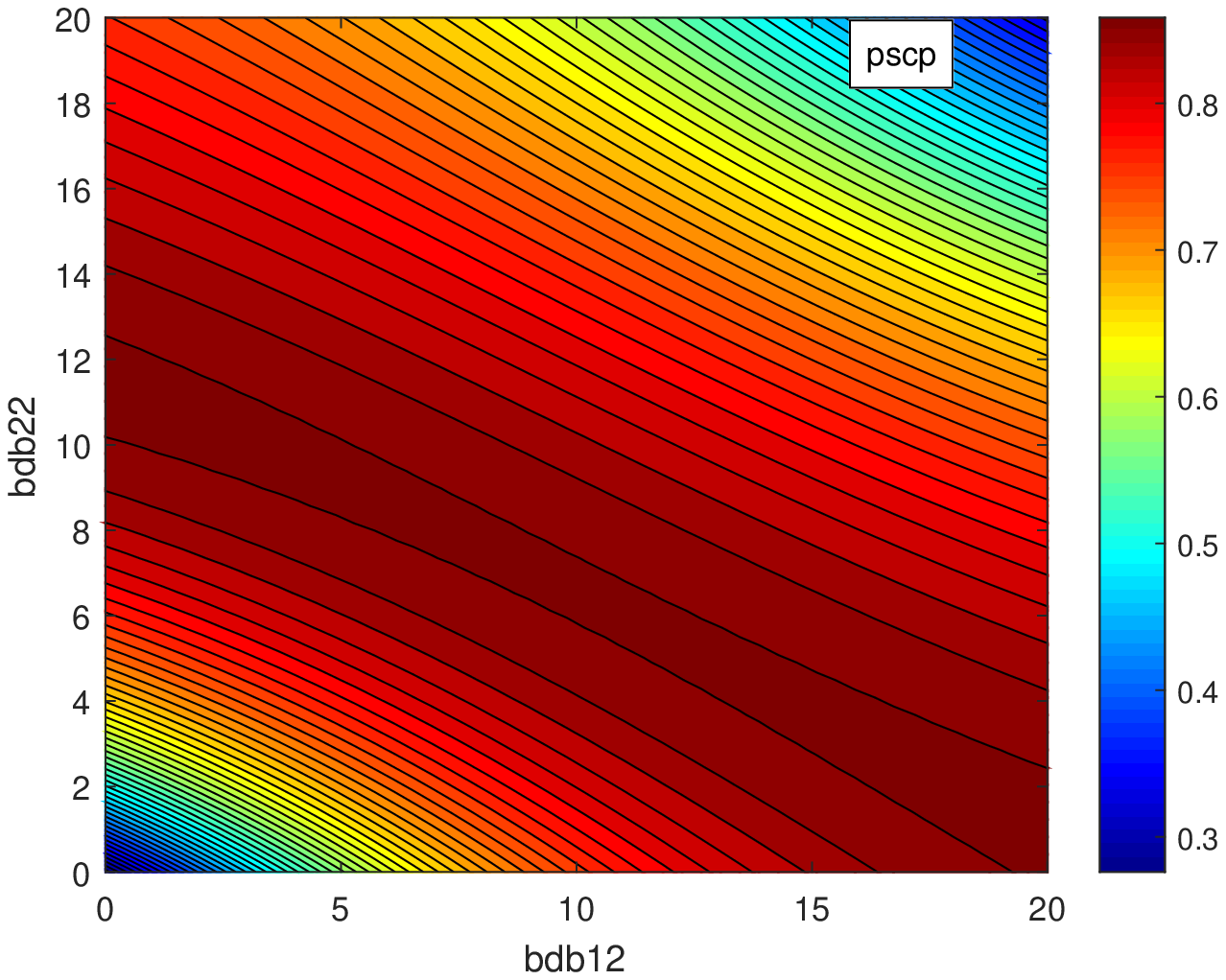}
		}
	\end{center}
	\vspace{-2mm}
	\caption{
		\ac{SCP} $\pcomp$
		of $2$ tier \ac{MEC}-enabled \ac{HetNet} with $2$ type user
		as a function of bias factor of a $1$-type user to $2$-tier \ac{MEC} server $\biasik{1}{2}$
		and bias factor of a $2$-type user to $2$-tier \ac{MEC} server $\biasik{2}{2}$.		
	}
	\vspace{0mm}
	\label{fig:scp22c}
\end{figure}

\begin{figure}
	\begin{center}   
		{ 
			\psfrag{bdb12}[tc][bc][0.7]{$\biasik{1}{2}$ [dB]}
			\psfrag{bdb22}[bc][tc][0.7]{$\biasik{2}{2}$ [dB]}
			\psfrag{pscp}[tc][tc][0.7]{$\pec$}
			\includegraphics[width=1.00\columnwidth]{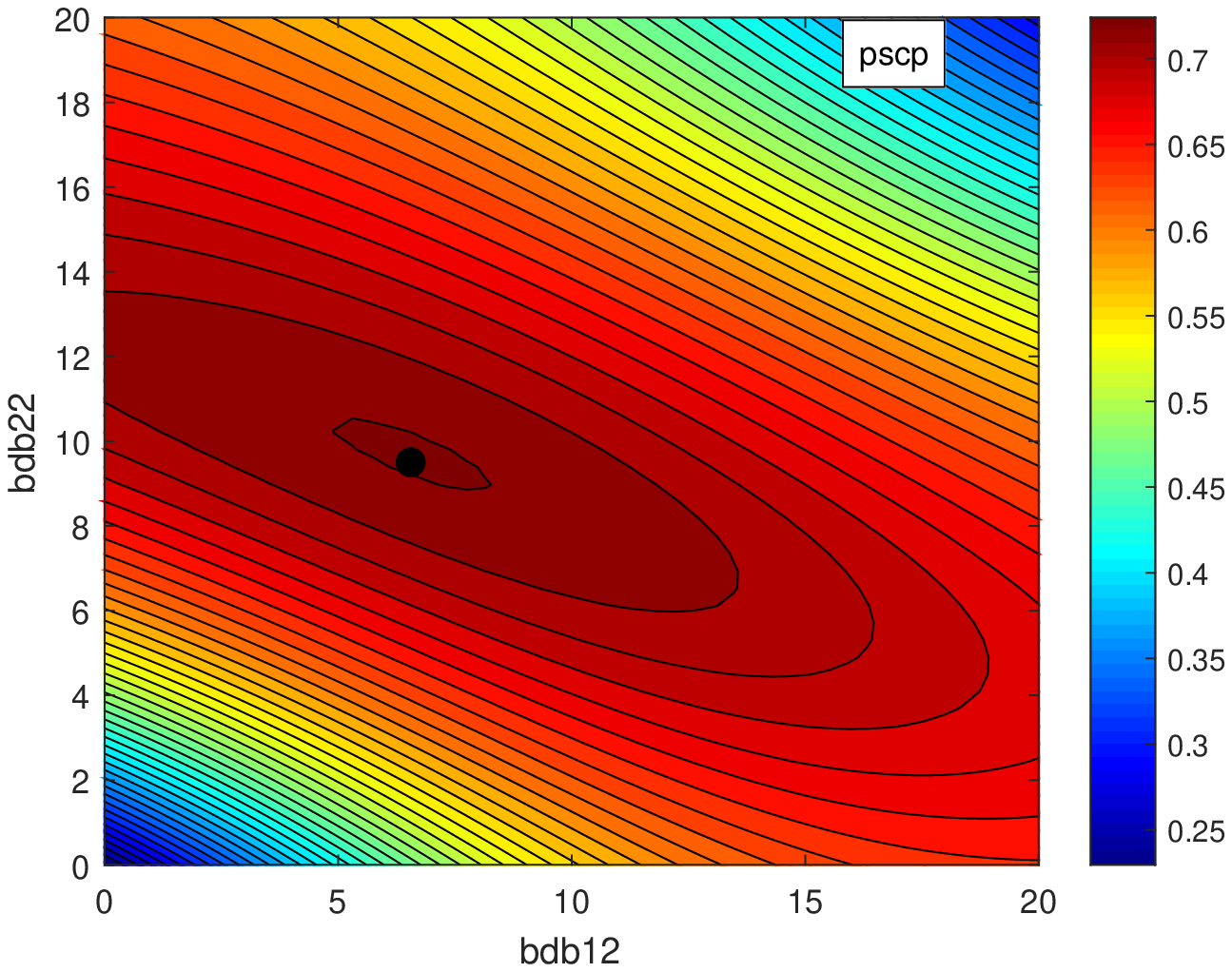}
		}
	\end{center}
	\vspace{-2mm}
	\caption{
		\ac{SECP} $\pec$
		of $2$ tier \ac{MEC}-enabled \ac{HetNet} with $2$ type user
		as a function of bias factor of a $1$-type user to $2$-tier \ac{MEC} server $\biasik{1}{2}$
		and bias factor of a $2$-type user to $2$-tier \ac{MEC} server $\biasik{2}{2}$.	
	}
	\vspace{0mm}
	\label{fig:secp22c}
\end{figure}

\begin{figure}
	\begin{center}   
		{ 
			\psfrag{bdb12}[tc][bc][0.7]{$\biasik{1}{2}$ [dB]}
			\psfrag{bdb32}[bc][tc][0.7]{$\biasik{3}{2}$ [dB]}
			\psfrag{pscp}[tc][tc][0.7]{$\pec$}
			\includegraphics[width=1.00\columnwidth]{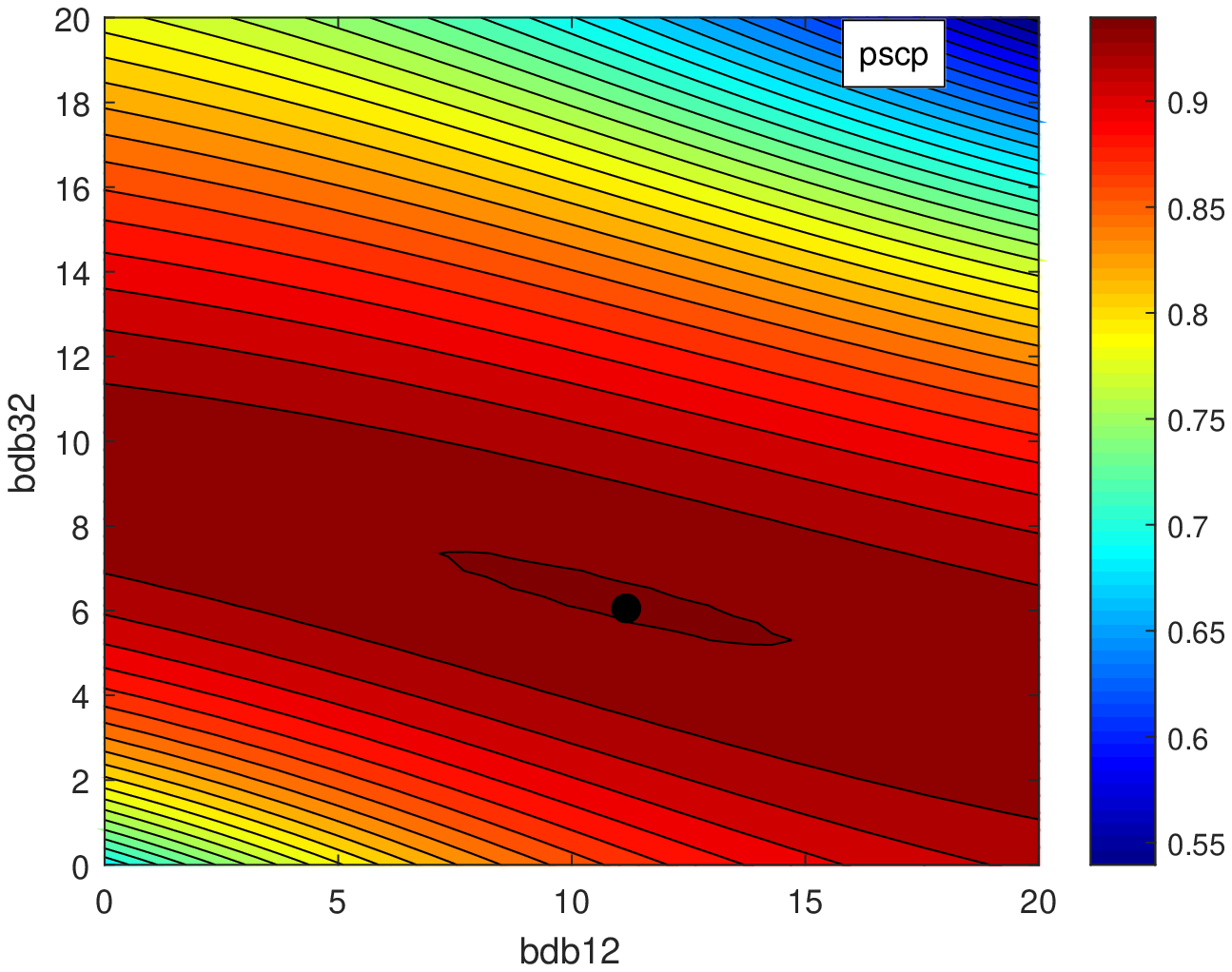}
		}
	\end{center}
	\vspace{-2mm}
	\caption{
		\ac{SECP} $\pec$
		of $2$ tier \ac{MEC}-enabled \ac{HetNet} with $3$ type user
		as a function of bias factor of a $1$-type user to $2$-tier \ac{MEC} server $\biasik{1}{2}$
		and bias factor of a $3$-type user to $2$-tier \ac{MEC} server $\biasik{3}{2}$.		
	}
	\vspace{0mm}
	\label{fig:secp23c}
\end{figure}

\blue{
Fig. \ref{fig:secp2tier} shows the contour of $\pec$ of $2$-tier network having $1$ types of users (i.e., $\SetI = \{1\}$ with $\dit{1}{(c)}=\{1\}$) as a function of $\biasik{1}{1}$ and $\biasik{1}{2}$ under the same environment of Fig. \ref{fig:secpscp21sim}.
It can be seen that the optimal bias factors are determined by the linear function of $\biasik{1}{1}$ and $\biasik{1}{2}$ because, according to \eqref{eq:association}, the user association is only adjuisted by the ratio between bias factors $\biasik{1}{1}/\biasik{1}{2}$.
}

\blue{
Fig. \ref{fig:secp3tier} shows the contour of $\pec$ of $3$-tier network having $1$-type user as a function of $\biasik{1}{2}$ and $\biasik{1}{3}$.
For this figure, $\biasik{1}{1}=10$dB, $\ttgi{1}=2\times10^{-3}$s, $\pwrmk{3}=23\text{dBm}$, $\lbdmk{3}=4.5\times10^{-5}$nodes/m$^{2}$, $\muk{3}=2$packet/slot, $\betawik{1}{3}=0.95$ for $q=\{$u,d$\}$, and other parameters are same as the Fig. \ref{fig:secp2tier}.
From Figs. \ref{fig:secp2tier} and \ref{fig:secp3tier}, we can find that $\biasoptik{1}{2}$ in Fig. \ref{fig:secp2tier} ($\biasoptik{1}{2}=8.1$dB when $\biasoptik{1}{1}=10$dB) is smaller than the one in Fig. \ref{fig:secp3tier} ($(\biasoptik{1}{2}, \biasoptik{1}{3})=(6.5, 9)\text{dB}$).
This implies that the computation tasks in Fig. \ref{fig:secp3tier} are distributed to the additional $3$-tier \ac{MEC} servers, which can achieve the better performance in terms of $\pec$.
It can be seen that the maximum \ac{SECP} in Fig. \ref{fig:secp3tier} (i.e., $\pec=0.95$) is bigger than the one in Fig. \ref{fig:secp2tier} (i.e., $\pec=0.93$).
Therefore, the \ac{SECP} can be improved by providing the additional tier of \ac{MEC} servers.
}

\blue{
Fig. \ref{fig:scp22c} and Fig. \ref{fig:secp22c} show the contour of $\pec$ and $\pcomp$, respectively, having $2$ types of users (i.e., $\SetI = \{1,2\}$ with $\dit{i}{(c)}=\{1,2\}$) as a function of $\biasik{1}{2}$ and $\biasik{2}{2}$.
First, from Fig. \ref{fig:scp22c}, we can see that $\pcomp$ decreases more by $\biasik{2}{2}$ than by $\biasik{1}{2}$.
This is because the variation of service time for large size tasks (i.e., $\dit{2}{(c)}$) is bigger than that for small size tasks (i.e., $\dit{1}{(c)}$), so $\pcomp$ becomes more sensitive by the arrival of $\dit{2}{(c)}$ size tasks.
}

\blue{
By comparing Figs. \ref{fig:scp22c} and \ref{fig:secp22c}, we can see that the optimal bias factors for $\pec$ and $\pcomp$ are different.
From Fig. \ref{fig:scp22c}, we can see that even when we offload all tasks of $1$-type user or $2$-type user to a $1$-tier \ac{MEC} server and no task to a $2$-tier \ac{MEC} server, i.e., $\biasoptik{1}{2} = 0$ or $\biasoptik{2}{2} = 0$, we can achieve the best performance in terms of $\pcomp$.  
However, it becomes different when we consider the \ac{SECP} as a performance metric.  
From Fig. \ref{fig:secp22c}, we can see that offloading certain amount of tasks to a $1$-tier \ac{MEC} server and a $2$-tier \ac{MEC} server, i.e., $\biasoptik{1}{2} > 0$ or $\biasoptik{2}{2} > 0$, can achieve the best performance in terms of $\pec$.
This is because in $\pec$, the communication performance is also considered, which can achieve low performance due to the longer link distance when all users associate to certain tier \ac{MEC} server.
Hence, the communication and computation performance needs to be considered together when we determine the bias factors for \ac{MEC} server association, which can be also seen for the case with $3$ types of users in Fig. \ref{fig:secp23c}.
}

\blue{
Fig. \ref{fig:secp23c} shows the contour of $\pec$ having $3$ types of users (i.e., $\SetI=\{1,2,3\}$ with $\dit{i}{(c)}=\{1, 2, 3\}$) as a function of the ratio of the bias factor $\biasik{1}{2}$ and $\biasik{3}{2}$ when $\biasik{2}{2}=10$dB. 
We can also see that $\pec$ decreases faster by $\biasik{3}{2}$ than $\biasik{1}{2}$ due to the large task size of a $3$-type users.
}

\subsection{\blue{SECP - Network Computation Capability}}

\begin{figure}
	\begin{center}   
		{ 
			\psfrag{secp}[bc][tc][0.8]{$\pec$}
			\psfrag{ratio}[tc][bc][0.7]{$\theta_{2}$}	
			\psfrag{secp-reference---------}[bl][bl][0.59]{$\ncck{2}=18*10^{-5}$}
			\psfrag{secp-ncc-low}[bl][bl][0.59]{$\ncck{2}=22.5*10^{-5}$}
			\psfrag{secp-ncc-high}[bl][bl][0.59]{$\ncck{2}=27*10^{-5}$}
			\includegraphics[width=1.00\columnwidth]{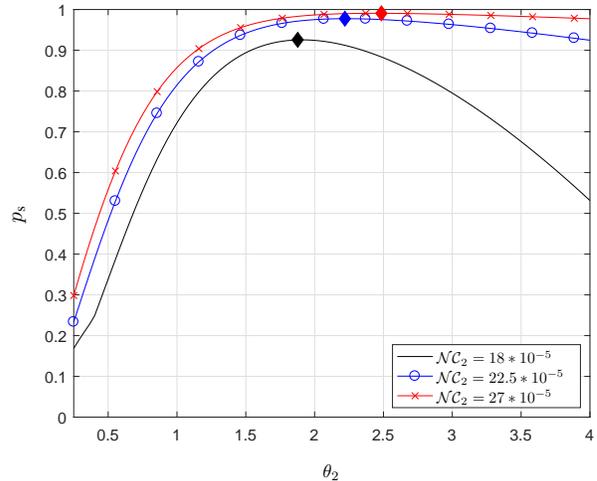}
		}
	\end{center}
	\vspace{-2mm}
	\caption{
		\ac{SECP} $\pec$
		of $2$ tier \ac{MEC}-enabled \ac{HetNet} with $2$ user type
		as a function of ratio factor $\theta_{2}$
		for different network computation capability $\ncck{2}$.		
	}
	\vspace{0mm}
	\label{fig:ncc12}
\end{figure}

\begin{figure}
	\begin{center}   
		{ 
			\psfrag{secp}[bc][tc][0.8]{$\pec$}
			\psfrag{bdb}[tc][bc][0.7]{$\biasik{2}{2}$ [dB]}	
			\psfrag{secp-reference----}[bl][bl][0.59]{$\theta_{1}=1$, $\mathbf{K}=1$}
			\psfrag{secp-ratio-low1}[bl][bl][0.59]{$\theta_{2}=0.8$, $\mathbf{K}=2$}
			\psfrag{secp-ratio-equal}[bl][bl][0.59]{$\theta_{2}=1$, $\mathbf{K}=2$}
			\psfrag{secp-ratio-high1}[bl][bl][0.59]{$\theta_{2}=1.2$, $\mathbf{K}=2$}
			\includegraphics[width=1.00\columnwidth]{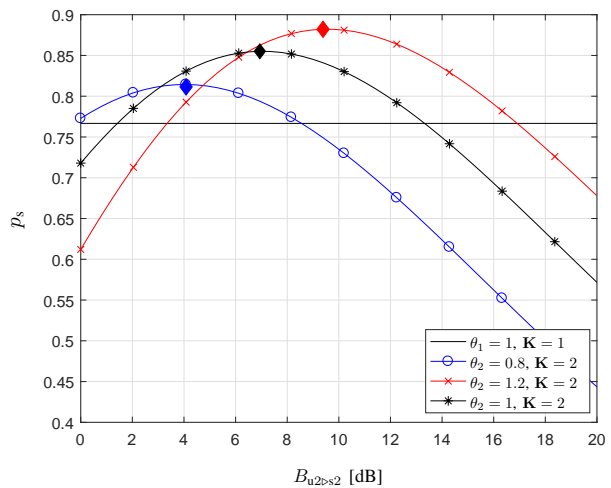}
		}
	\end{center}
	\vspace{-2mm}
	\caption{
		\ac{SECP} $\pec$
		of $1$ and $2$ tier \ac{MEC}-enabled \ac{HetNet} with $2$ user type
		as a function of bias factor of a $2$-type user to a $2$-tier \ac{MEC} server $\biasik{2}{2}$
		for different ratio factor $\theta_{2}$.		
	}
	\vspace{0mm}
	\label{fig:ncc22}
\end{figure}

\blue{
In this subsection, we have also provided additional design insights by introducing the concept of the network computation capability. The network computation capability $\ncck{\mathbf{K}}$ is defined as the total computing capability, available in the network. Hence, it is proportional to the number of \ac{MEC} servers and the CPU frequency of each \ac{MEC} servers, given by
\begin{align} \label{eq:defncck}
\ncck{\mathbf{K}}=\sum_{j\in\SetK}\lbdmk{j}^{\theta_{j}}\muk{j}^{\theta_{k}}=\sum_{j\in\SetK}\left(\frac{1}{\theta_{j}}\lbdmk{j}\right)\left(\theta_{j}\muk{j}\right)
\end{align}
where $\lbdmk{j}$ and $\muk{j}$ are the spatial density and computing capability (i.e., service rate) of $j$-tier \ac{MEC} servers, respectively.
In this framework, the number of \ac{MEC} servers and the CPU frequency of \ac{MEC} servers are proportional to the spatial density of \ac{MEC} servers, and the computing capability of \ac{MEC} servers, respectively.
Hence, $\ncck{\mathbf{K}}$ can be presented by $\lbdmk{j}$ and $\muk{j}$.
The $\theta_{j}$ is the ratio factor of the $j$-tier \ac{MEC} servers.
From \eqref{eq:defncck}, we can see that $\ncck{\mathbf{K}}$ remains unchanged even when $\theta_{j}$ increases or decreases.
For instance, if $\theta_{j}$ increases, the $\muk{j}$ increases, but $\lbdmk{j}$ decreases.
From Figs. \ref{fig:ncc12} and \ref{fig:ncc22}, we discuss how to deploy \ac{MEC} servers when $\ncck{\mathbf{K}}$ is given. To this end, we have adjusted the density of \ac{MEC} servers as $\lbdmk{j}/\theta_{j}$ and the CPU frequency as $\theta_{j}\muk{j}$ to see their impact when $\ncck{\mathbf{K}}$ is fixed.
}


\blue{
Fig. \ref{fig:ncc12} shows $\pec$ as a function of $\theta_{2}$ for different network computation capabilities $\ncck{\mathbf{K}}$.
For this figure, $\theta_{1}=1$, $\biasik{2}{2}=15$dB, and other parameters are same as Fig. \ref{fig:secpscp22mu}.
We can see that as $\theta_{2}$ becomes larger, $\pec$ increases to the optimal points.
For small $\theta_{2}$, in spite of the short link distance due to the high \ac{MEC} server density in $2$-tier $\lbdmk{2}$, low computing capability of $2$-tier \ac{MEC} server $\muk{2}$ cause the degradation of $\pec$.
As $\theta_{2}$ increases, $\muk{2}$ increase, which can achieve higher $\pec$ even though $\lbdmk{2}$ gradually decreases.
However, for excessively high $\theta_{2}$, the larger communication latency due to the low $\lbdmk{2}$ leads to the decrease in $\pec$ in spite of the high $\muk{2}$.
}

\blue{
Moreover, it can be seen that as $\ncck{\mathbf{K}}$ increases, $\pec$ becomes higher.
Consequently, as $\ncck{\mathbf{K}}$ increases, the optimal $\theta_{2}$ maximizing $\pec$ also becomes higher.
This implies that when $\ncck{\mathbf{K}}$ is large, the optimal $\theta_{2}$ is determined to increase $\muk{2}$ rather than $\lbdmk{2}$.
Hence, when the total computation capability deployed in the network is fixed, increasing the computing capability and reducing the number of \ac{MEC} servers are more beneficial to enhance the \ac{SECP}.
}

\blue{
Fig. \ref{fig:ncc22} presents $\pec$ as a function of $\biasik{2}{2}$ for the single tier network and $2$ tier network with different ratio factor $\theta_{2}$.
For the single tier network, the total \ac{MEC} server density $\lbdm$ is same as $\lbdm$ for the $2$ tier network, and the computing capability of \ac{MEC} server $\muk{1}$ is determined so that $\ncck{1}$ for the single tier network has same value with $\ncck{2}$ for $2$ tier network.
Other parameters for single tier network are same as the $1$-tier network in the $2$ tier \ac{MEC}-enabled \ac{HetNet}
}

\blue{
It can be seen that $\pec$ for multi-tier cases can be larger than $\pec$ for single-tier case.
Specifically, for $\biasik{2}{2}$ in the range from $1.4$dB to $13.2$dB when $\theta_{2}=1$, $\pec$ for $2$ tier network is larger than $0.77$, which is the value of $\pec$ for single tier network.
This implies that extending the single-tier to multi-tier networks can achieve higher $\pec$ without increasing the $\ncck{\mathbf{K}}$ when $\biasik{2}{2}$ is adjusted by considering both the computation and the communication performance.
Therefore, when the network computation capability is fixed, the \ac{SECP} in multi-tier networks can be higher than the \ac{SECP} in single-tier networks by adjusting the association bias factor.
}

\blue{
Moreover, we can see that $\biasoptik{2}{2}$ for $\theta_{2}=1.2$ is larger than the one for $\theta_{2}=0.8$.
As $\theta_{2}$ increases, the computing capabilities of $2$-tier \ac{MEC} servers $\muk{2}$ become larger and the density of $2$-tier servers $\lbdmk{2}$ becomes smaller.
As a result, the users who offload to the $2$-tier \ac{MEC} servers can have higher $\pec$ because the total arrivals to $2$-tier server decrease due to the low $\lbdmk{2}$, and high $\muk{2}$.
Hence, $\biasoptik{2}{2}$ becomes larger to distribute the tasks to the $2$-tier \ac{MEC} servers.
}

\section{Conclusions} \label{sec:conclusions}

In this paper, we propose the \ac{MEC}-enabled \ac{HetNet}, composed of the multi-type users with different computation task sizes and the multi-tier \ac{MEC} servers with different computing capacities.
\blue{We derive the \ac{SECP} by analyzing the distribution of total latency in \ac{MEC}-enabled \ac{HetNet}.
With the consideration of both the computing time in \ac{MEC} server and the transmission time depending on the target coverage probability, the closed form of \ac{SECP} for special cases, and the approximated \ac{SECP} for general case are obtained.}
We then evaluate the effects of bias factors in association and network parameters on the \ac{SECP}. 

\blue{Our results provide some insights on the design of \ac{MEC}-enabled \ac{HetNet}.
Specifically, 
1) the \ac{MEC}-enabled \ac{HetNet} has different optimal association bias factors from conventional ones, which consider the communication performance only or the computation performance only, and the optimal bias factor is located between the optimal bias factors obtained in terms of computing or communication performance only,
2) offloading more computation tasks to the high-capable \ac{MEC} servers generally shows better \ac{SECP} unless those servers are heavy-loaded,
3) when the computation tasks of the network becomes large, 
it is better to distribute the arrival of tasks to the low-capable \ac{MEC} servers instead of offloading the most of all tasks to the high-capable \ac{MEC} servers,
4) when the total computation capability deployed in the network is fixed, increasing the computing capability and reducing the number of \ac{MEC} servers are more beneficial to enhance the SECP, and
5) when the network computation capability is fixed, the SECP in \emph{multi-tier} networks can be higher than the SECP in \emph{single-tier} networks.}
\begin{appendix}
\subsection{\blue{Proof of Lemma~\ref{coro:cdf2123}}} \label{app:corocdf2123}
\blue{
To derive the \ac{SECP}, the \ac{cdf} of the waiting time at the \ac{MEC} servers (i.e., M/G/1 queue) is required, which has no closed form to the best of our knowledge. 
Hence, we approximate the distribution of the waiting time to the Gamma distribution after calculating the Gamma distribution-related parameters such as $\beta_{k,1}$ and $\beta_{k,2}$ in \eqref{eq:b1}, respectively.
According to \cite{kle:75}, the \ac{cdf} of the waiting time $F\left(t\right)$, obtained by performing the inverse Laplace transform of \eqref{eq:applappk}, is presented by
\begin{align} \label{eq:appwpdf}
F\left(t\right)=
1-\rhok{k}+\rhok{k}\fgbb{t}
\end{align}
where $\rhok{k}$ is the utilization factor and $\fgbb{t}$ is the \ac{cdf} of the waiting time for tasks, which are not immediately served upon arrival.
We approximate $\fgbb{t}$ to the Gamma distribution, which is also applied in \cite{menhenzeptra:06}.
By using the Takacs Recursion Formula in \cite{kle:75}, the mean waiting time and the mean square of waiting time for $k$-tier \ac{MEC} server is obtained by
\begin{align} \label{eq:appgammatakacs1}
\ex{\tquk{k}}
=\frac{\sum_{i\in\SetI}\lbdaik{i}{k}\ex{\tsvik{i}{k}^{2}}}{2\left(1-\rhok{k}\right)}
\end{align}
\begin{align} \label{eq:appgammatakacs2}
\ex{\tquk{k}^{2}}
=2\ex{\tquk{k}}^{2}
+\frac{\sum_{i\in\SetI}\lbdaik{i}{k}\ex{\tsvik{i}{k}^{3}}}{3\left(1-\rhok{k}\right)}
\end{align}
where $\ex{\!\tsvik{i}{k}^{2}\!}$ and $\ex{\!\tsvik{i}{k}^{3}\!}$ are defined, respectively, by
\begin{align} \label{eq:appgammatsv2}
\ex{\!\tsvik{i}{k}^{2}\!}
\!=\!\!\int_{0}^{\infty}\!\!\!\!t^{2}\!f_{\tsvik{i}{k}}\!\!\left(t\right)\!dt
\!=\!\frac{\dit{i}{(c)}\!\left(\dit{i}{(c)}+1\right)}{\muk{k}^{2}}
\end{align}
\begin{align} \label{eq:appgammatsv3}
\ex{\!\tsvik{i}{k}^{3}\!}
\!=\!\!\int_{0}^{\infty}\!\!\!\!t^{3}\!f_{\tsvik{i}{k}}\!\!\left(t\right)\!dt
\!=\!\frac{\dit{i}{(c)}\!\left(\dit{i}{(c)}+1\right)\!\left(\dit{i}{(c)}+2\right)}{\muk{k}^{3}}.
\end{align}
By matching the \eqref{eq:appgammatakacs1} and \eqref{eq:appgammatakacs2} to a mean and variance of Gamma distribution, the Gamma distribution-related parameters $\beta_{k,1}$ and $\beta_{k,2}$ are obtained by \eqref{eq:b1}.
Applying the \eqref{eq:b1}, the approximated \ac{cdf} of the waiting time $\hat{F}\left(t\right)$ is given by
\begin{align} \label{eq:appapxwpdf}
\hat{F}\left(t\right)=
1-\rhok{k}+\rhok{k}\frac{\gamma\left(\beta_{k,1},\beta_{k,2}t\right)}{\Gamma\left(\beta_{k,1}\right)}
\end{align}
for $t\geq0$ where $\gamma\left(\cdot,\cdot\right)$ and $\Gamma\left(\cdot,\cdot\right)$ are the lower and upper incomplete gamma function, respectively.
Substituting the part of the equation \eqref{eq:pcompikint} into \eqref{eq:appapxwpdf}, $\pecikh{i}{k}$ is obtained by
\begin{align} \label{eq:apppecikh_1}
\pecikh{i}{k}=&
\int_{0}^{\infty}
\left[1-\rhok{k}+\rhok{k}\hat{F}\left(T_{\text{th},i}-r\right)\right]f_{\tsvik{i}{k}}\left(r\right)dr \nonumber \\
=&
\left(1-\rhok{k}\right)F_{\gamma,1}+\rhok{k}F_{\gamma,2}
\end{align}
where $T_{\text{th},i}$ is $\ttg-\sum_{q\in\SetQ}\tcmwik{i}{k}$ for $q=\{\text{u}, \text{d}\}$. In \eqref{eq:apppecikh_1}, $F_{\gamma,1}$ and $F_{\gamma,2}$ are given, respectively, by
\begin{align} \label{eq:apppecikh_1_1}
F_{\gamma,1}\!\!=\!\!
\int_{0}^{T_{\text{th},i}}\!\!\frac{\muk{k}^{\dit{i}{(c)}}r^{\dit{i}{(c)}\!-1}\!\exp\{\!-\muk{k}r\!\}}{\left(\dit{i}{(c)}-1\right)!}dr
\!=\!
\frac{\gamma(\dit{i}{(c)},\muk{k}T_{\text{th},i})}{\left(\dit{i}{(c)}-1\right)!}
\end{align}
\begin{align} \label{eq:apppecikh_1_2}
F_{\gamma,2}\!=\!\!&
\int_{0}^{T_{\text{th},i}}\!\!\frac{\gamma\!\left(\beta_{k,1},\beta_{k,2}\!\left(T_{\text{th},i}\!-\!r\right)\right)}{\Gamma\left(\beta_{k,1}\right)}\frac{\muk{k}^{\dit{i}{(c)}}r^{\dit{i}{(c)}-1}\exp\{\!-\muk{k}r\}}{\left(\dit{i}{(c)}-1\right)!}dr \nonumber \\
=&
\frac{\gamma\!\left(\beta_{k,1},\beta_{k,2}T_{\text{th},i}\right)}{\Gamma\left(\beta_{k,1}\right)}
-\frac{\!\exp\left\{-\beta_{k,2}T_{\text{th},i}\right\}}{\beta_{k,2}^{-\beta_{k,1}}\Gamma\left(\beta_{k,1}\right)} \nonumber \\
&\times\sum_{n=0}^{\dit{i}{(c)}-1}\!\frac{\muk{k}^{n}}{n!}\!\int_{0}^{T_{\text{th},i}}\!\frac{\exp\{\left(\beta_{k,2}\!-\!\muk{k}\right)r\}r^{k}}{\left(T_{\text{th},i}-r\right)^{1-\beta_{k,1}}}dr.
\end{align}
From the equation in \cite[eq. (3.383)]{GraRyz:B07}, $F_{\gamma,2}$ is provided as the closed-form equations. By substituting the \eqref{eq:apppecikh_1_1} and \eqref{eq:apppecikh_1_2} into \eqref{eq:apppecikh_1}, the approximated \ac{SECP} becomes \eqref{eq:cdf2123}. }
\end{appendix}
\bibliographystyle{IEEEtran}
\bibliography{Bibtex/StringDefinitions,Bibtex/IEEEabrv,Bibtex/mybib_cw}
\end{document}